\documentclass[onecolumn]{emulateapj}

\def\lesssim{\mathrel{\hbox{\rlap{\hbox{\lower4pt\hbox{$\sim$}}}\hbox{$<$}}}}
\def\gtrsim{\mathrel{\hbox{\rlap{\hbox{\lower4pt\hbox{$\sim$}}}\hbox{$>$}}}}

\def\hmag{H_{\rm mag}}
\def\hrho{H_\rho}
\def\hrad{H_{\rm rad}}

\def\B{\begin{equation}}
\def\E{\end{equation}}

\def\B{\begin{equation}}
\def\E{\end{equation}}

\font\gkvec=cmmib10
\def\bxi{\hbox{{\gkvec\char24}}} 

\usepackage{graphicx}

\slugcomment{}
\shorttitle{Radiation MHD Instabilities}
\shortauthors{Tao \& Blaes}

\begin{document}

\title{Local Radiation MHD Instabilities in Magnetically Stratified Media}

\author{Ted Tao and Omer Blaes}
\affil{Department of Physics, University of California,
Santa Barbara CA 93106}

\begin{abstract}
We study local radiation magnetohydrodynamic instabilities in static, optically
thick, vertically stratified media with constant flux mean opacity.  We
include the effects of vertical gradients in a horizontal
background magnetic field.  Assuming rapid radiative diffusion, we use the zero
gas pressure limit as an entry point for investigating the coupling between the
photon bubble instability and the Parker instability.
Apart from factors that depend on wavenumber
orientation, the Parker instability exists for wavelengths longer
than a characteristic wavelength $\lambda_{\rm tran}$, while photon bubbles
exist for wavelengths shorter
than $\lambda_{\rm tran}$.  The growth rate in the Parker regime is
independent of the orientation of the horizontal component of the wavenumber
when radiative diffusion is rapid, but the range of Parker-like wavenumbers is
extended if there exists strong horizontal shear between field
lines (i.e. horizontal wavenumber perpendicular to the magnetic field).
Finite gas pressure introduces an additional short wavelength limit to the
Parker-like behavior, and also limits the growth rate of the photon bubble
instability to a constant value at short wavelengths.  We also consider the
effects of differential rotation with accretion disk applications in mind.
Our results may explain why photon bubbles have not
yet been observed in recent stratified shearing box accretion disk simulations.
Photon bubbles may physically exist in simulations with high radiation to gas
pressure ratios, but higher spatial resolution will be needed to resolve the
asymptotically growing unstable wavelengths. 
\end{abstract}


%%%%%%%%%%%%%%%%%
\section{Introduction}

Considerable attention has been devoted recently to the study of radiation
pressure driven instabilities in magnetized media, which have generally
come to be known as photon bubble instabilities.  This name is
particularly appropriate to the buoyant long wavelength instabilities studied
by \citet{aro92} in the context of accreting X-ray pulsars.  \citet{gam98}
found that an analogous instability is likely to exist in
radiation-dominated accretion disks.  \citet{bla01,bla03} extended Gammie's
analysis to finite gas sound speeds and short wavelengths to show that the
instability in this regime is a radiative amplification of magnetosonic modes.
At least in media where Thomson scattering is the dominant opacity, the
slow mode always has the fastest growth rate.  \citet{beg01} derived
a fully nonlinear periodic shock train solution to the equations of radiation
MHD, which he suggested would be the nonlinear outcome of the short wavelength
photon bubble instability.  This was later confirmed by numerical simulation
\citep{tur05}.
\citet{beg06a} has also discovered a nonlinear wave solution
in the long wavelength, buoyancy regime of the photon bubble instability.

Applications of the short wavelength photon bubble instability to accretion
disks have been explored extensively by \citet{beg02,beg06b}.  Perhaps
the most significant implication is that the instability might permit highly
super-Eddington luminosities from accretion disks that would still be
geometrically thin in terms of their vertical mass distribution.
The physics of the short wavelength photon bubble instability has also
been extended to situations where other forms of diffusive energy transport
exist.  In particular, work has been done on ``neutrino bubble instabilities''
in proto-neutron stars \citep{soc05} and ``Coulomb bubble instabilities'' in
systems with anisotropic thermal conduction due to the presence of a magnetic
field \citep{soc08}.

All of this work on photon bubble and related instabilities has assumed a
uniform background magnetic field, but gradients in the magnetic field can
also drive instabilities.  In particular, a medium may be
vulnerable to magnetic interchange and undulatory Parker instabilities
\citep{tse60,new61,par66,par67} if magnetic pressure gradients contribute 
significant support against gravity. Vertically stratified shearing box
simulations indicate that such instabilities, not photon bubbles, appear to
dominate the large scale dynamics of the surface layers of accretion disks
\citep{bla07}, though it is conceivable that photon bubbles would be
relevant on smaller length scales that are unresolved in the existing
simulations.

If both the equilibrium density and magnetic pressure decrease outward, then both the Parker and photon bubble unstable modes reduce to slow magnetosonic modes in the short wavelength limit.  It is therefore likely that these two instabilities have nontrivial couplings in equilibria with both magnetic and radiation pressure gradients, and it is important to study this coupling in order to understand under what regimes and length scales
each instability is likely to operate.  This is the purpose of the present paper:  we investigate local radiation MHD instabilities in equilibria that are supported against gravity by both radiation and magnetic pressure
gradients. 

We begin in section 2 by stating the basic radiation MHD equations and
assumptions that we employ in our analysis. The perturbation equations are very
complicated to analyze when we include both radiation and magnetic field
physics, and much of the rest of the paper is devoted to various levels of
approximation. The situation is simplest when one neglects gas pressure
completely, as in Gammie's (1998) original analysis of the photon bubble
instability.  We employ that approach in section 3, and find a single unstable
mode at all wavelengths.  There is a characteristic transition wavelength
$\lambda_{\rm tran}$
above which the mode is Parker-like and below which the mode becomes the
standard photon bubble instability, although the actual transition wavelength
depends on the orientation of the wave vector of the perturbations.
In section 4, we discuss the effects of
finite gas pressure.  Because much of the recent interest in photon
bubble instabilities lies in accretion disks, we include the effects of
differential rotation in section 5 and then discuss applications
of our results to accretion disks in section 6.  We finish by summarizing
our findings in section 7. In the appendices we briefly
show that our perturbation equations recover the basic properties of magnetic
buoyancy instabilities derived by previous authors and show that, in contrast
to the case of adiabatic perturbations \citep{new61}, the most rapidly
growing Parker modes do not require large horizontal shear between field lines
when radiation diffusion is rapid.  We also present a more detailed derivation
of our instability analysis with differential rotation.

\section{Equations}

\citet{bla03} derived simplified versions of the general radiation MHD equations
of \citet{sto92}, and we continue to use the same equations here, but with some
additional assumptions.
We assume that the gas and radiation temperatures are locked together and
that the flux mean opacity $\kappa$ is a constant (as would be
true for Thomson scattering).  The gravitational field ${\bf g}=-g\hat{\bf z}$,
with $g>0$ possibly being a function of height $z$, but independent of time
as we neglect the effects of self-gravity.  The equations then become

\begin{equation}
{\partial\rho\over\partial t}+{\nabla}\cdot(\rho{\bf v})=0,
\label{eqcont}
\end{equation}

\begin{equation}
\rho\left({\partial{\bf v}\over\partial t}+{\bf v}\cdot\nabla{\bf v}\right)=
-\nabla p+\rho{\bf g}+{1\over4\pi}{\bf B}\cdot\nabla{\bf B}-{1\over8\pi}
\nabla B^2+{\kappa\rho\over c}{\bf F},
\label{eqmom}
\end{equation}

\begin{equation}
{\partial\over\partial t}(e+E)+{\bf v}\cdot\nabla(e+E)+\left({4\over3}E+
\gamma e\right)\nabla\cdot{\bf v}=-\nabla\cdot{\bf F},
\label{eqenergy}
\end{equation}

\begin{equation}
{\bf F}=-{c\over 3\kappa\rho}\nabla E,
\end{equation}

\begin{equation}
{\partial{\bf B}\over\partial t}=\nabla\times({\bf v}\times{\bf B}),
\end{equation}

\noindent and

\begin{equation}
e={p\over\gamma-1},\,\,\,\,\,
p={\rho k_{\rm B}T\over\mu},\,\,\,\,\,{\rm and}\,\,\,\,\,
E=aT^4.
\label{eqstate}
\end{equation}

Here $\rho$ is the fluid mass density, ${\bf v}$ is the fluid velocity,
$p$ is the gas pressure, $e$ is the gas internal energy density, $T$ is the
temperature, $E$ is the radiation energy density, ${\bf F}$ is the
radiation flux, and ${\bf B}$ is the magnetic field.  Other symbols have
their usual meanings: $c$ is the speed of light, $a$ is the radiation density
constant, $k_{\rm B}$ is Boltzmann's constant,  $\mu$ is the mean particle
mass in the gas, and $\gamma$ is the ratio of specific heats in the gas.

\subsection{Equilibrium}

We assume a static, vertically stratified, horizontally homogeneous
equilibrium with purely horizontal magnetic field ${\bf B}=B(z)\hat{\bf y}$.
The equilibrium radiation flux is ${\bf F}=F\hat{\bf z}$.
The only differential equations that the equilibrium
must satisfy are those of hydrostatic equilibrium,
\begin{equation}
{d\over dz}\left(p+{E\over3}+{B^2\over8\pi}\right)=-\rho g,
\label{eqhydrostatic}
\end{equation}
radiative equilibrium,
\begin{equation}
{dF\over dz}=0,
\label{eqradiative}
\end{equation}
and radiative diffusion
\begin{equation}
F=-{c\over3\kappa\rho}{dE\over dz}.
\label{eqraddiff}
\end{equation}

\subsection{Perturbations}

We linearize equations (\ref{eqcont})-(\ref{eqstate}) about the general
equilibrium, assuming a $(t,x,y)$-dependence of $\exp[i(k_xx+k_yy-\omega t)]$.
We do not make any assumptions about the $z$-dependence of the perturbations
as our background is a function of $z$ and we in particular wish to include
the effects of the background magnetic gradient to the maximum possible extent.
Eliminating magnetic and radiation flux perturbations, we obtain five coupled
ordinary differential equations:

\begin{equation}
-i\omega\tilde{\delta\rho}+ik_x\delta v_x+ik_y\delta v_y+{d\delta v_z\over dz}
-{\delta v_z\over\hrho}=0,
\label{eqcontpert}
\end{equation}

\begin{equation}
i(\omega^2-k_x^2v_{\rm A}^2-k_y^2v_{\rm A}^2)\delta v_x-
i{k_x\omega\over\rho}\delta P
-k_xv_{\rm A}^2{d\delta v_z\over dz}
+{k_xv_{\rm A}^2\over2\hmag}\delta v_z=0,
\end{equation}

\begin{equation}
i\omega^2\delta v_y-i{k_y\omega\over\rho}\delta P+{k_yv_{\rm A}^2
\over2\hmag}\delta v_z=0,
\end{equation}

\begin{eqnarray}
& &iv_{\rm A}^2{d^2\delta v_z\over dz^2}-{3iv_{\rm A}^2\over2\hmag}
{d\delta v_z\over dz}+i\left[\omega^2-k_y^2v_{\rm A}^2+{v_{\rm A}^2\over
(\hmag^\prime)^2}\right]\delta v_z-k_xv_{\rm A}^2{d\delta v_x\over dz}+
{k_xv_{\rm A}^2\over \hmag}\delta v_x\cr
& &-{\omega\over\rho}{d\delta P\over dz}-\omega g\tilde{\delta\rho}=0,
\label{eqmomzpert}
\end{eqnarray}
and
\begin{eqnarray}
& &{4cE\over3\kappa\rho}{d^2\tilde{\delta T}\over dz^2}
-\left(8F-{4cE\over3\kappa\rho \hrho}\right)
{d\tilde{\delta T}\over dz}
+\left[{i\omega\over\Gamma_3-1}\left(p+{4E\over3}\right)-
{4cE\over3\kappa\rho}(k_x^2+k_y^2)\right]
\tilde{\delta T}\cr
& &+F{d\tilde{\delta\rho}\over dz}
+{i\omega\rho\over\Gamma_3-1}(c_{\rm i}^2-c_{\rm t}^2)\tilde{\delta\rho}-
{\rho c_{\rm t}^2N^2\over g(\Gamma_3-1)}\delta v_z=0.
\label{eqenergypert}
\end{eqnarray}
Here $v_{\rm A}=B/\sqrt{4\pi\rho}$ is the Alfv\'en speed, $\tilde{\delta\rho}\equiv\delta\rho/\rho$,
$\tilde{\delta T}\equiv\delta T/T$, and $\delta P$ is the total thermal pressure
perturbation,
\begin{equation}
\delta P\equiv\delta p+{\delta E\over3}=\rho c_{\rm i}^2\tilde{\delta\rho}+
\left(p+{4E\over3}\right)\tilde{\delta T}.
\label{eqpressurepert}
\end{equation}
The quantities $\hrho(z)$, $\hmag(z)$, and
$\hmag^\prime(z)$ are measures of the local density and
magnetic pressure scale heights,
\begin{equation}
\hrho\equiv-\left({d\ln\rho\over dz}\right)^{-1}\,\,\,\,\,\,\,\,\,\,
\hmag\equiv-\left({d\ln B^2\over dz}\right)^{-1}\,\,\,\,\,\,\,\,\,\,
\hmag^\prime\equiv\left[{1\over2}{d^2\ln B^2\over dz^2}+{1\over2}
\left({d\ln B^2\over dz}\right)^2\right]^{-1/2}.
\end{equation}
We implicitly assume that $\hrho$ and $\hmag$ are non-negative, i.e. that
density and magnetic field do not increase outward.  This is the regime
of most relevance for the outer layers of stellar envelopes and the uppermost
layers of accretion disks.
The quantity $N^2$ is the square of the local Brunt-V\"ais\"al\"a frequency
in the gas-radiation mixture,
\begin{equation}
N^2=g\left[{1\over\Gamma_1}{d\over dz}\ln\left(p+{E\over3}\right)-
{d\ln\rho\over dz}\right],
\end{equation}
$c_{\rm i}=(p/\rho)^{1/2}$ is the isothermal sound speed in the gas,
$c_{\rm t}=[\Gamma_1(p+E/3)/\rho]^{1/2}$ is the total adiabatic sound speed
in the gas plus radiation mixture,
and $\Gamma_1$ and $\Gamma_3$ are generalized adiabatic exponents \citep{cha67},
\begin{equation}
\Gamma_1={16E^2+60(\gamma-1)Ee+9\gamma(\gamma-1)e^2\over9(e+4E)(p+E/3)}
\,\,\,\,\,{\rm and}\,\,\,\,\,\Gamma_3={16E+3\gamma e\over3(e+4E)}.
\end{equation}

\section{The Zero Gas Pressure Limit}
\label{sectionpzero}

The perturbation equations presented above are complicated and cannot be
readily combined into a single ordinary differential equation.  To make
progress, we first consider short wavelengths where radiative diffusion
is rapid, and follow \citet{gam98} by adopting the limit of zero gas
pressure.  After all, we are particularly interested in environments where
magnetic and radiation pressures dominate greatly over gas pressure.
The mathematical advantage of this approximation is that there
is then no slow magnetosonic wave limit and the photon bubble
instability appears at lowest order in the short wavelength limit.  (For
nonzero gas pressure, this analysis would only be valid for photon bubble
wavelengths longer than a finite turnover wavelength; see
section~\ref{sec:finitegas} below.)
At the same time, the minimum unstable wavelength for the Parker instability
also vanishes in the rapid diffusion limit if the gas pressure is zero. Hence the zero gas pressure limit
enables us to explore the coupled photon bubble and Parker problem in the
short wavelength WKB limit at lowest order.  We shall find that the shortest
wavelengths are always in the photon bubble regime, but if magnetic support
dominates radiation support in the equilibrium, then there is a WKB transition
wavelength longward of which the Parker instability becomes manifest.

Proceeding with a WKB ansatz that the $z$-dependence of all
perturbations is $\exp(ik_zz)$, we obtain a seventh order dispersion relation.
Taking the infinitely high wavenumber limit with $\omega\sim k^2$ gives the
usual damped diffusion mode,
\begin{equation}
\omega=-i{ck^2\over3\kappa\rho},
\label{eqdiffmode}
\end{equation}
where $k^2\equiv k_x^2+k_y^2+k_z^2$.
Doing the same thing with $\omega\sim k$ gives the two fast magnetosonic
modes,
\begin{equation}
\omega^2=k^2v_{\rm A}^2,
\label{eqfastmode}
\end{equation}
and the two Alfv\'en modes,
\begin{equation}
\omega^2=k_y^2v_{\rm A}^2.
\label{eqalfvenmode}
\end{equation}

The remaining two modes describe the coupled Parker and photon bubble
instabilities.  Assuming that $\omega$ grows no faster than $k^{1/2}$ as
$k\rightarrow\infty$, the lowest order terms in the dispersion relation
become
\begin{equation}
\omega^2k^4+i\omega\left({4\kappa E\over3c}\right)k_y^2k^2
+\left[{k^2(k_x^2+k_y^2)v_{\rm A}^2\over4\hmag^2}+(k^2k_x^2+k^2k_y^2+
2k_z^2k_y^2){\kappa F\over2\hmag c}-
ik_zk_y^2k^2{\kappa F\over c}
\right]=0.
\label{eqdisp1}
\end{equation}
Clearly as $k\rightarrow\infty$, the lowest order solution is given by
\begin{equation}
\omega^2=i{k_y^2k_z\over k^2}{\kappa F\over c},
\label{eqomegapbi}
\end{equation}
identical to the photon bubble dispersion relation in this regime
\citep{gam98}.

On the other hand, if magnetic pressure gradients dominate radiation pressure
gradients in supporting the medium against gravity, the last term in equation
(\ref{eqdisp1}) can be small except at extremely short wavelengths.  Neglecting
this term, and also the linear term in $\omega$ for the moment, the dispersion
relation gives for large magnetic pressure gradients
\begin{equation}
\omega^2=-{(k_x^2+k_y^2)v_{\rm A}^2\over4k^2\hmag^2}\simeq -{(k_x^2+k_y^2)g\over
2k^2\hmag},
\label{eqomegaparker}
\end{equation}
which is the short wavelength Parker instability growth rate in this regime.
This result agrees
with the $c_{\rm i}=0$ and $k\rightarrow\infty$ (with $\omega\sim k^0$) limit
of the Parker dispersion relation derived by \citet{bla07} in their appendix.
Their result assumed an isothermal medium, implying $F=0$.  However, we show
in Appendix~\ref{parkerappendix} that it is also valid for more general
equilibria in which
radiation and magnetic pressures dominate gas pressure, and radiative diffusion
is rapid.  [See equation (\ref{eqgilmancizero}).]

Note that the pure Parker instability is a genuine exponentially growing
instability in all space, whereas the pure photon bubble instability is an overstability of
traveling waves.  Nevertheless, the different wavenumber dependencies of
equations (\ref{eqomegapbi}) and (\ref{eqomegaparker}) indicate that a
transition between photon bubble and Parker behavior will occur
at a characteristic vertical wavenumber $k_{\rm tran}$ given by
\begin{equation}
k_{\rm tran}\equiv{2\pi\over\lambda_{\rm tran}}
\equiv{v_{\rm A}^2c\over4\hmag^2\kappa F},
\end{equation}
provided at least that $k_x$ is not too large compared to $k_y$.  Note that
the photon bubble growth rate always declines with increasing $k_x$ at fixed
$k$.  The photon bubble instability grows fastest when the perturbation
gradients lie entirely in the plane defined by gravity and the magnetic field.
The Parker growth rate is relatively unaffected by $k_x$ in this zero gas
pressure regime.

The corresponding transition wavelength can also be written in a more
physically transparent fashion as
\begin{equation}
\lambda_{\rm tran}=4\pi H_{\rm mag}\left({dP_{\rm rad}/dz\over dP_{\rm mag}/dz}
\right),
\end{equation}
where $P_{\rm rad}=E/3$ is the radiation pressure and $P_{\rm mag}=B^2/(8\pi)$
is the magnetic pressure.  As magnetic pressure gradient forces become stronger
compared to radiation pressure forces, $\lambda_{\rm tran}$ decreases, extending
the long-wavelength range of the Parker instability at the expense of the
short-wavelength range of the photon bubble instability.

For WKB to make sense with $k<k_{\rm tran}$, we require
$k_{\rm tran}\hmag\gg1$.  This implies that
magnetic support must dominate radiation pressure support in the equilibrium.
If this is not true, then there is no WKB-accessible range of wavelengths
where the Parker instability will be present (except for very short radial
wavelengths: $k_x^2/k_y^2\gg(k_{\rm tran}\hmag)^{-1}$).  Assuming
$k_{\rm tran}\hmag\gg1$, the middle term in the square brackets in
equation (\ref{eqdisp1}) is negligible, leaving us with
\begin{equation}
\omega^2+i{k_y^2\over k^2}{4E\kappa\over3c}\omega
+{(k_x^2+k_y^2)v_{\rm A}^2\over4\hmag^2k^2}-i{k_y^2k_z\over k^2}
{\kappa F\over c}=0.
\label{eqdispparkerpbi}
\end{equation}
If damping (the linear term in $\omega$) is negligible, the solution to
this dispersion relation is
\begin{eqnarray}
\omega&=&\pm{k_yv_{\rm A}\over2^{3/2}k\hmag}\Biggl\{
\left[\left((1+k_x^2/k_y^2)^2+{k_z^2\over k_{\rm tran}^2}\right)^{1/2}-
(1+k_x^2/k_y^2)\right]^{1/2}\cr
& &+i\left[\left((1+k_x^2/k_y^2)^2+{k_z^2\over k_{\rm tran}^2}\right)^{1/2}+
(1+k_x^2/k_y^2)\right]^{1/2}
\Biggr\}.
\end{eqnarray}
The positive root is unstable:  photon bubbles are recovered for
$k_z\gg(1+k_x^2/k_y^2)k_{\rm tran}$, and Parker is recovered for
$k_z\ll(1+k_x^2/k_y^2)k_{\rm tran}$.

To understand the role of the damping term, it is helpful to introduce the
dimensionless quantity
\begin{equation}
b\equiv{4E\kappa\hmag\over3cv_{\rm A}}=3\left(\frac{c_{\rm r}}{c}\right)\left(\frac{c_{\rm r}}{v_{\rm A}}\right)
\kappa\rho\hmag,
\end{equation}
where $c_{\rm r}=(4E/9\rho)^{1/2}$ is the effective radiation sound speed.
It is also convenient to scale the angular frequency of the mode with $v_{\rm A}/\hmag$,
\begin{equation}
\tilde{\omega}\equiv{\omega \hmag\over v_{\rm A}}.
\end{equation}
Then the solution of the dispersion relation (\ref{eqdispparkerpbi}) may
be written as
\begin{equation}
\tilde{\omega}=-i{k_y^2b\over2k^2}\pm{k_y\over2k}
\left[-\left(1+{k_x^2\over k_y^2}+{k_y^2b^2\over k^2}\right)
+i{k_z\over k_{\rm tran}}\right]^{1/2}
\end{equation}
or
\begin{eqnarray}
\tilde{\omega}&=&-i{k_y^2b\over2k^2}\pm{k_y\over2^{3/2}k}\times\cr
              & &\Biggl\{\left[\left((1+k_x^2/k_y^2+k_y^2b^2/k^2)^2+{k_z^2\over
                 k_{\rm tran}^2}\right)^{1/2}-(1+k_x^2/k_y^2+k_y^2b^2/k^2)
                 \right]^{1/2}\cr
              & &+i\left[\left((1+k_x^2/k_y^2+k_y^2b^2/k^2)^2+{k_z^2\over
                 k_{\rm tran}^2}\right)^{1/2}+(1+k_x^2/k_y^2+k_y^2b^2/k^2)
                 \right]^{1/2}
                 \Biggr\}.
\label{eqzerogasgrowth}
\end{eqnarray}
Note that the mode with the upper (plus) sign is always unstable, in that
the imaginary part of the frequency is always positive.  The other root is
always damped.  The unstable mode again transitions smoothly from Parker
for $k_z\ll(1+k_x^2/k_y^2+k_y^2b^2/k^2)k_{\rm tran}$ to photon bubble
for $k_z\gg(1+k_x^2/k_y^2+k_y^2b^2/k^2)k_{\rm tran}$.  The instability
always becomes Parker-like for wave vectors in the $x-z$ plane, i.e.
$k_y\rightarrow0$.

Damping affects the Parker regime most.  For
$k_z\ll(1+k_x^2/k_y^2+k_y^2b^2/k^2) k_{\rm tran}$, we have
\begin{equation}
\tilde{\omega}\simeq-i{k_y^2b\over2k^2}\pm i{k_y\over2k}\left(1+
{k_x^2\over k_y^2}+{k_y^2b^2\over k^2}\right)^{1/2}.
\end{equation}
Unless $k_y\rightarrow0$, the growth rate of the unstable root is greatly
reduced when damping is large ($b\gg1$):
$\tilde\omega\simeq i(1+k_x^2/k_y^2)/(4b)$.  On the other hand, photon bubbles
are relatively unaffected by damping provided $k_z/k_{\rm tran}\gg b^2$.
If instead $b^2\gg k_z/k_{\rm tran}\gg1$, then the real and imaginary parts
of the unstable photon bubble frequency are markedly reduced:
\begin{equation}
\omega\simeq{4E\kappa\over3c}\left({k_z\over4k_{\rm tran}b^2}
+i{k_z^2k^2\over16k_{\rm tran}^2k_y^2b^4}\right).
\end{equation}

\begin{figure}
\epsscale{1.0}
%\plotone{gammasimp.eps}
\plotone{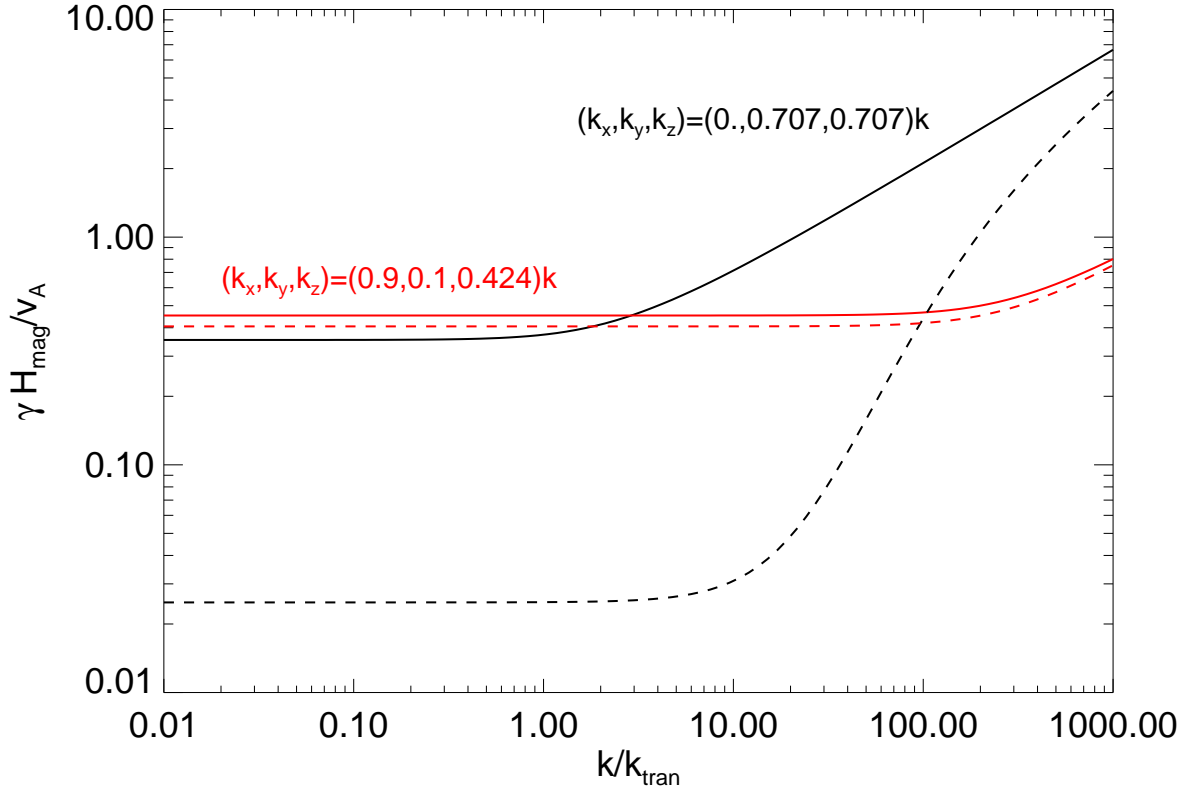}
\caption{Scaled unstable mode growth rate $\gamma\equiv {\rm Im}(\omega)$
as a function of the dimensionless
wavenumber $k/k_{\rm tran}$.  The different colors are for different wavenumber
orientations:  $\hat{\bf k}=(0,0.707,0.707)$ (black) and
$\hat{\bf k}=(0.9,0.1,0.424)$ (red).  Solid curves have no damping ($b=0$),
while dashed curves have $b=10$.}
\label{fig:gammasimp}
\end{figure}

Figure \ref{fig:gammasimp} illustrates these behaviors.  The black curves show
cases with $k_x=0$ and zero damping ($b=0$, solid curve) and finite damping
($b=10$, dashed).  Here the mode transitions from
constant (Parker) growth rate at low wavenumbers to a (photon bubble)
growth rate that increases as $k^{1/2}$ for high wavenumbers.  The transition
occurs at $k\sim k_{\rm tran}$ with zero damping, and for $k$ somewhere
between 10 and 100~$k_{\rm tran}$ for $b=10$.  The latter is consistent with
the transition wavenumber being given by $k_z\sim k_y^2b^2k_{\rm tran}/k^2$,
implying $k\sim (k_y/k)b^2k_{\rm tran}=70.7k_{\rm tran}$.  For this wavenumber
orientation, damping affects the Parker instability severely.  The photon
bubble growth rate is much less affected, however:  little damping occurs
for $k_z>b^2k_{\rm tran}$ or $k>(k/k_z)b^2k_{\rm tran}=140k_{\rm tran}$.
The red pair of curves have the same damping parameters
but now have wavenumbers oriented mostly in the $x-z$ plane.
The transition to photon bubble behavior now occurs above a much higher
wavenumber $k\sim(k_x^2/k_y^2)(k/k_z)k_{\rm tran}\sim200k_{\rm tran}$.
The Parker instability is insensitive to damping for this wavenumber
orientation. 

\section{Finite Gas Pressure Effects}
\label{sec:finitegas}

We now consider the effects of nonzero gas pressure on the coupled photon
bubble/Parker instability.  Finite gas pressure stabilizes the Parker
instability at infinitely short wavelengths.  Then radiative diffusion is so
fast that $\delta T\approx0$, and instability only occurs for wavenumbers
satisfying
\begin{equation}
|k_y|<{(k_x^2+k_y^2)^{1/2}\over k}k_{\rm P},
\label{parker}
\end{equation}
where
\begin{equation}
k_P\equiv\left({g\over2H_{\rm mag}c_{\rm i}^2}\right)^{1/2}.
\label{eqkparker}
\end{equation}
[See equation (\ref{eqkp2gilman}) in the Appendix.] If the magnetic
pressure gradient is the dominant force in supporting the medium against
gravity, then $k_{\rm P}$ reduces to
$\approx g/(c_{\rm i}v_{\rm A})=v_{\rm A}/(2\hmag c_{\rm i})$.
In this case equation (\ref{parker}) can be written as
$k_yv_{\rm A}<g/c_{\rm i}$ (ignoring wavenumber factors), which reveals
its physical origin.  Buoyancy is fundamentally what drives the Parker
instability.  Stability therefore requires that the magnetic tension
not have time to restore the perturbed fluid elements on a buoyant rise time
in the gas in the ambient gravitational field (which is largely determined by
magnetic pressure gradients).  Note, however, that the growth rate of the
fastest mode of the instability is $\sim v_{\rm A}/H_{\rm mag}$, and occurs at
a wavenumber much smaller than the marginally stable wavenumber $k_{\rm P}$
[equation (\ref{eqk0cizero}) in the Appendix].

In a medium that is only supported against gravity by gas and radiation
pressure (no magnetic field gradients), finite sound speed affects the photon
bubble instability by causing the growth rate to reach a fixed value once the
wavenumber is high enough for the mode to become a slow magnetosonic mode.
Provided that gas and radiation temperatures are thermally locked (as we are
assuming throughout this paper), this turnover wavenumber is given by
\begin{equation}
k_{\rm T}={\kappa F\over cc_{\rm i}^2}\left(1+{3p\over4E}\right).
\end{equation}
For wavenumbers greater than $k_{\rm T}$, the instability growth rate
asymptotes to a constant value $\sim k_{\rm T}c_{\rm i}$ when the magnetic
pressure is much larger than the gas pressure $p$ \citep{bla03}.

Previous studies of the short wavelength photon bubble instability assumed
a uniform equilibrium magnetic field, which therefore provides no support
against gravity \citep{gam98,bla01,bla03}.  In this case, the turnover
wavenumber $k_{\rm T}$ is approximately $g/c_{\rm i}^2$, i.e. the reciprocal
of the gas pressure scale height \citep{bla03}.  When magnetic pressure
gradients support the equilibrium, $g/c_{\rm i}^2$ can be significantly
larger than the true turnover wavenumber, implying that the asymptotic
short wavelengths of the photon bubble instability (which have the highest
growth rates) can be more easily resolved in numerical simulations than one
might have expected from just the gas pressure scale height.  On the other
hand, the asymptotic growth rate can also be significantly smaller than
the radiation pressure supported equilibrium estimate $g/c_{\rm i}$ when
magnetic gradients support the equilibrium.

We can write the three wavenumber scales in the problem as
\begin{equation}
k_{\rm tran}=\frac{\hrad}{3\hmag^2}
\left(\frac{v_{\rm A}}{c_{\rm r}}\right)^2,
\label{ktran}
\end{equation}
\begin{equation}
k_{\rm T}=
\left(1+{3c_{\rm r}^2\over c_{\rm i}^2}\right){1\over4H_{\rm rad}}
\label{kt}
\end{equation}
and
\begin{equation}
k_{\rm P}=
\left\{{1\over2H_{\rm mag}}\left[\left(1+{3c_{\rm r}^2\over c_{\rm i}^2}\right)
{1\over4H_{\rm rad}}+{v_{\rm A}^2\over2c_{\rm i}^2H_{\rm mag}}+
{1\over H_{\rho}}\right]\right\}^{1/2},
\label{kp}
\end{equation}
where $\hrad\equiv-(d\ln E/dz)^{-1}$ is the radiation pressure scale height,
$k_{\rm T}$ and $k_{\rm P}$ are the photon bubble turnover and Parker cutoff
wavenumbers, respectively.  We emphasize that these expressions are only valid
in the limit of rapid radiative diffusion.
To study our problem around these wavenumber regimes with
WKB methods, we need $kH\gg1$, where $k$ is any of $k_{\rm tran}$, $k_{\rm T}$
or $k_{\rm P}$ and $H$ is any of the scale heights.

\begin{figure}
%\plotone{f2new.eps}
\plotone{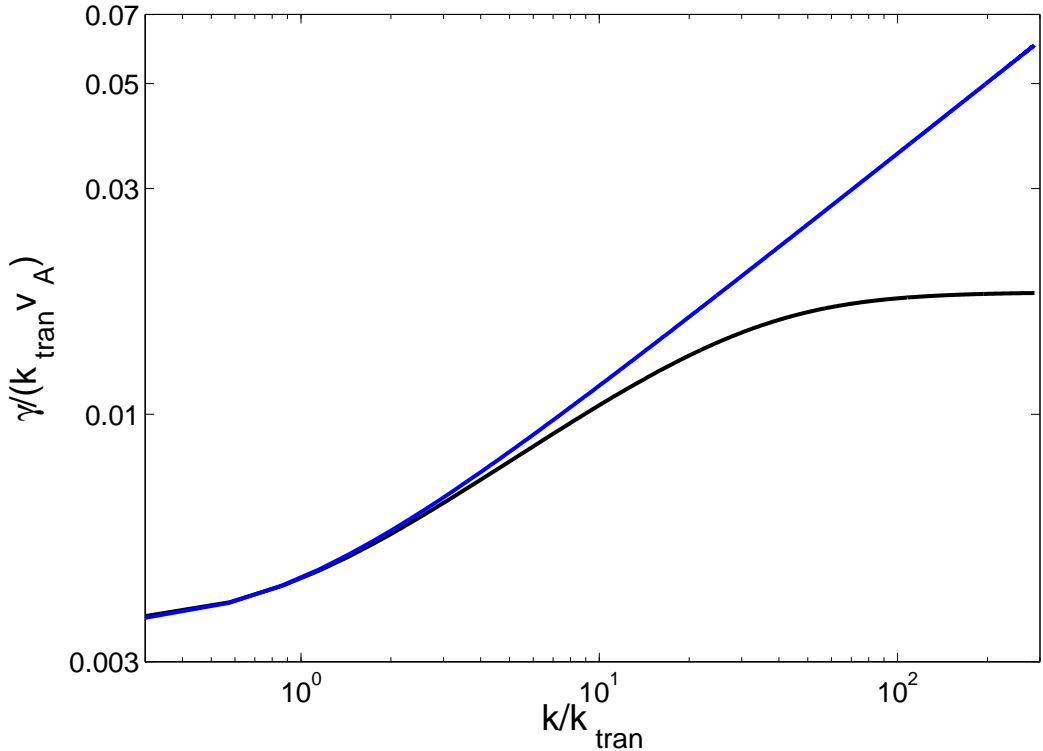}
\caption{Instability growth rate $\gamma\equiv{\rm Im}(\omega)$ (black curve)
in units of $k_{\rm tran}v_{\rm A}$ versus total wavenumber in units of
$k_{\rm tran}$, for a wavenumber orientation $k_x:k_y:k_z=\sin(\pi/12):1:1$.
Here we set all the scale heights in the calculation to be equal:
$H_{\rm mag}=H_\rho=H_{\rm rad}$.  The sound speeds, Alfv\'en speed, and
damping parameter satisfy $c_{\rm i}/v_{\rm A}=0.01$,
$c_{\rm r}/v_{\rm A}=0.1$,
and $b=0.01$.  This implies $k_{\rm tran}H\simeq33$,
$k_{\rm P}/k_{\rm tran}\simeq50$, and $k_{\rm T}/k_{\rm tran}\simeq75$.
The blue curve in this figure shows the growth rate from the analytic zero gas
sound speed expression (\ref{eqzerogasgrowth}).
}
\label{fig:finitegas}
\end{figure}

As we showed in section~\ref{sectionpzero}, our $c_{\rm i}=0$ WKB dispersion
relation (\ref{eqdispparkerpbi}) agrees with previously
published photon bubble and Parker instability growth rates. This suggests
that taking the $z$-dependence of all perturbations to be ${\rm exp}(ik_{z}z)$
without combining the perturbation equations first may prove useful even in
the finite but small (compared to magnetic and radiation pressures) gas
pressure regime. We solved the resulting dispersion relation numerically.
Figure \ref{fig:finitegas} shows results for an illustrative set of parameters.
The photon bubble growth rate asymptotes to a
finite value beyond the turnover wavenumber $k_{\rm T}$ because the gas
pressure is no longer exactly zero.  

In both Figures \ref{fig:gammasimp} and \ref{fig:finitegas}, note that there is
no sign of photon bubbles for low wavenumber ($k\ll k_{\rm tran}$) where
diffusion will eventually be slow. In particular, our low wavenumber growth rate
agrees with that of the Parker instability, but not with the \cite{aro92} photon
bubble result. Thus if magnetic gradients are large enough that $k_{\rm tran}$
corresponds to wavelengths in the rapid diffusion regime, then the slow
diffusion
version of the photon bubble instability may not exist.  However, a non-WKB
analysis is necessary to verify this conclusively.

It is also possible in principle for the maximum Parker growth rate
to be higher than that of the photon bubble instability.  Analytically,
this occurs when the photon bubble growth rate obtained by \cite{bla03} (their
equation 93) exceeds the Parker growth rate stated in equation
(\ref{eqgammamaxgilman}). In the limit that the medium is magnetically
supported, and the gas pressure is much smaller than radiation and magnetic
pressures, this comparison means that Parker dominates if
$v_{\rm A}/H_{\rm mag}\gg (c_{\rm r}/H_{\rm rad})(c_{\rm r}/c_{\rm i})$.
Provided the scale heights are comparable, this is equivalent to the condition
that $k_{\rm P}<k_{\rm tran}$, so that the Parker growth rate cuts off and
decreases down to the asymptotic photon bubble growth rate.  In this case the
Parker instability transitions into the photon bubble instability when $k$
exceeds $k_{\rm P}$, which is {\it less} than the transition wavenumber
$k_{\rm tran}$ for zero gas pressure.  For finite gas pressure, the Parker
growth rate can exceed the photon bubble growth rate even without this
ordering of the characteristic wavenumbers.  We will later present a finite
gas pressure example of this from a numerical simulation of
accretion disk vertical structure in section~\ref{sec:accretiondisks} below.

\section{Effects of Rotation and Shear}
\label{secrotationshear}

Motivated by the fact that both Parker and photon bubble instabilities are
driven by vertical gradients, and by possible applications to stellar
envelopes and atmospheres, our analysis so far has assumed a static equilibrium.
However, there has been considerable recent interest in photon bubble physics
in radiation dominated accretion disks.  Such flows are differentially
rotating, and before one can apply our results to such flows, one has to
account for the effects of rotation and shear.  Such effects were first
examined by \citet{shu74} for the Parker instability, and by
\citet{bla01} in the case of the photon bubble instability restricted
to axisymmetric perturbations.

In the spirit of the short wavelength WKB approximation that we have
been using, we restrict attention to a local, corotating patch of a
differentially
rotating disk and use the fluid equations describing a shearing box
\citep{gol65,haw95} with coordinate axes $(x,y,z)$ aligned with the
radial, azimuthal, and vertical directions, respectively.
Appendix \ref{secrotationderivation} presents a
detailed analysis of the coupled Parker-photon bubble problem, again assuming
negligible gas pressure as we did in section~\ref{sectionpzero} above.
Provided $k_{\rm tran}H_{\rm mag}\gg1$, the azimuthal component $\xi_y$ of
the small amplitude Lagrangian displacement of perturbations satisfies
the differential equation
\begin{equation}
{\partial^2\xi_y\over\partial t^2}+{k_y^2\over k^2}{4E\kappa\over3c}
{\partial\xi_y\over\partial t}-{(k_x^2+k_y^2)v_{\rm A}^2\over4
H_{\rm mag}^2k^2}\xi_y+i{k_y^2k_z\over k^2}{\kappa F\over c}\xi_y=0.
\label{eqxidiffeq}
\end{equation}
In the absence of differential rotation, all the coefficients of this
equation would be independent of time.  We could then assume a complex
exponential time-dependence $\xi_y\propto\exp(-i\omega t)$ and recover
the static equilibrium dispersion relation (\ref{eqdispparkerpbi}) that
we derived above.  As we discuss in more detail in
Appendix~\ref{secrotationderivation}, Coriolis forces are negligible
provided the WKB approximation is valid.

However, the shear due to the differential rotation of the equilibrium flow
has a more important effect.  This shear means that the radial wavenumber
$k_x$ depends on time:  $k_x(t)=k_{x0}+q\Omega tk_y$,
where $k_{x0}$ is the initial value of $k_x$, $q$ is the shear parameter
(3/2 for a Keplerian accretion disk),
and $\Omega$ is the angular velocity of the local patch of disk.
The exponential growth rates of the static equilibrium analysis will therefore
only be valid so long as these growth rates are
faster than the time it takes the radial wavenumber to change significantly
in time.  For a magnetically supported medium, this implies that
$(k_{x0}^2/k^2)(k^2_\perp/k_y^2)\gg q^2$ for the Parker regime, and
$(k_{x0}^2/k^2)(|k_z|/k_{\rm tran})\gg q^2$ for the photon bubble regime.
The latter is easier to satisfy than the former, as the transition from
Parker to photon bubble occurs when $|k_z|/k_{\rm tran}>k^2_\perp/k_y^2$,
assuming negligible damping.  On the other hand, if radiation pressure
supports the equilibrium, then shear will not affect the exponential
growth of photon bubbles provided $(k_{x0}^2/k^2)|k_z|H_{\rm rad}\gg q^2$,
which is essentially equivalent to the WKB condition.  Short wavelength
photon bubbles are therefore generally less affected by shear than Parker
modes.

\begin{figure}
\plotone{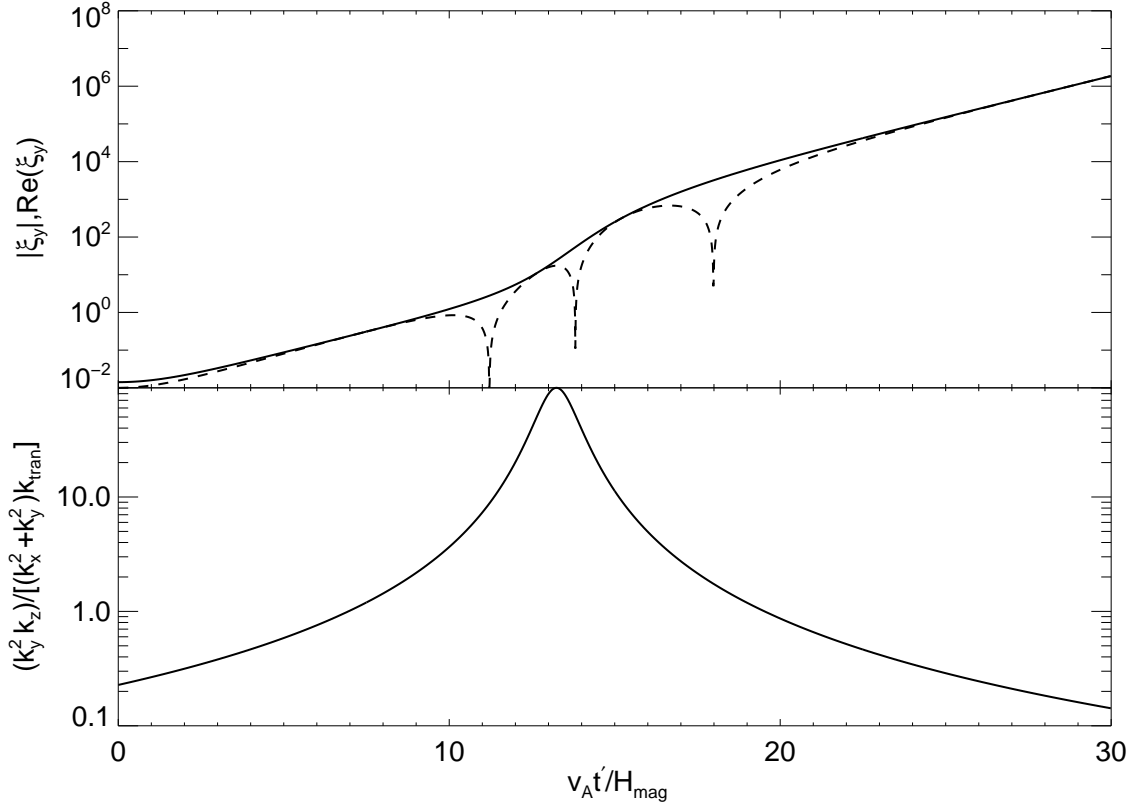}
\caption{Upper plot: modulus (solid curve) and real part (dashed curve) of
the azimuthal component of the Lagrangian displacement, in arbitrary units,
as a function of time in a shearing box with negligible gas pressure.
This particular case assumed dimensionless parameter values of $b=0$
(i.e. negligible radiative damping), $\Omega H_{\rm mag}/v_{\rm A}=1$, $q=3/2$,
$k_z/k_{\rm tran}=90$, and $(k_{x0},k_y,k_z)=(-0.993,0.05,0.107)k$.
Lower plot: the dimensionless ratio
$k_y^2k_z/(k_\perp^2k_{\rm tran})$ as a function of time for the same
parameters.  The instability is in the Parker regime when this ratio is less
than of order unity, and in the photon bubble regime otherwise.
}
\label{fig:shearbox}
\end{figure}

Note that if these conditions are not satisfied and/or the instabilities fail
to reach the nonlinear regime before the waves start to shear, then the
character of the unstable driving can change with time.  For example, a
strongly leading nonaxisymmetric perturbation (e.g. $k_y>0$ and
$k_{x0}$ large and negative)
could start in the Parker regime, then flip to the photon bubble regime
as the wave swings from leading to trailing ($k_x\sim0$), and then flip back
to the Parker regime as the perturbation becomes strongly trailing ($k_x$
large and positive).  We illustrate this linear regime behavior in
Figure~\ref{fig:shearbox} which depicts a numerical solution to
equation~(\ref{eqxidiffeq}).  Nonaxisymmetric photon bubbles inevitably
enter the Parker regime unless they become nonlinear first as the wavevector
shears to be strongly trailing.  But again, photon bubbles whose vertical
wavelengths are short enough should be able to grow much faster than the
shear rate.

Axisymmetric ($k_y=0$) perturbations are immune to these shear effects, but
the photon bubble growth rate vanishes for such perturbations.  Photon bubbles
necessarily require wave vectors which are neither entirely perpendicular
or along the equilibrium magnetic field \citep{bla03}.  Because we are
considering equilibria with vertical magnetic gradients, we restricted
consideration to equilibrium magnetic fields that are entirely in the
azimuthal ($y$) direction.  These are generally the dominant magnetic
field components even in a turbulent accretion disk because of the differential
rotation.  But other components are present, especially in the surface
layers if large-scale Parker modes are able to grow \citep{bla07}.
Axisymmetric photon bubbles on an equilibrium with magnetic field components
in all directions have growth rates that are proportional to the square of
the ratio of the non-azimuthal field components to the total magnetic field
\citep{gam98,bla01,bla03}, and hence have slower growth rates than
nonaxisymmetric photon bubbles in a medium dominated by azimuthal fields.

Our discussion in this section has assumed negligible sound speed $c_{\rm i}$
in the gas alone.  As we will see in the next section, radiation MHD simulations
of local patches of accretion disks generally produce conditions in which
$k_{\rm T}>k_{\rm P}>k_{\rm tran}$.  Hence the main thing that finite gas
sound speed introduces is a limit $\sim k_{\rm T} c_{\rm i}$ to the short
wavelength photon bubble growth rate for $k\gtrsim k_{\rm T}$.  In that
very short wavelength regime, photon bubbles continue to be immune to the
shear rate.  For a radiation pressure supported medium, this merely requires
$\Omega H_{\rm rad}/c_{\rm i}\gg q$, which is clearly satisfied because
$\Omega\sim c_{\rm r}/H_{\rm rad}$.  For a magnetically supported medium,
this requires instead that $c_{\rm r}^2/(c_{\rm i}v_{\rm A})\gg q$ provided
the radiation pressure and magnetic pressure scale heights are comparable.
This in turn is equivalent to the condition that $k_{\rm T}\gg k_{\rm tran}$,
which we find to be generally true.

To summarize, in a medium supported by radiation pressure, rotation and shear
do not significantly affect the photon bubble regime provided we are considering
vertical wavelengths that are short enough for the WKB approximation to be valid
($|k_z|H\gg1$).  In a magnetically supported medium, photon bubbles grow
quickly compared to the shear rate provided the vertical wavelength is much
shorter than the transition wavelength $\lambda_{\rm tran}$.  Shear is
more important near or longer than the transition wavelength; i.e. in the
Parker regime, modes can be more affected by shear, unless they are
associated with very short radial wavelengths.

\section{Applications to Astrophysical Accretion Disks}
\label{sec:accretiondisks}

In this section we investigate the relevance of photon bubble and Parker
instabilities to the surface layers of high luminosity accretion disks, using
recent stratified shearing box, radiation-MHD simulations as a guide
to expected local conditions.  In particular, we use data from simulations
0528a of \citet{kro07} and \citet{bla07}, and 1112a and 0519b of \citet{hir09}.
The simulations represent local patches of accretion disks at 30
(1112a and 0519b) and 150 (0528a) gravitational radii around a 6.62 solar mass
black hole, and have time-averaged, volume-integrated radiation to gas
pressure ratios of approximately 1 (0528a), 7 (1112a), and 70 (0519b).
We time and horizontally average the fluid variables in each simulation
in order to have an approximate vertically stratified equilibrium in which
to explore the instabilities that we have calculated above.
All time averages of the simulation
data neglect the first ten orbits during which the magnetorotational
instability was still developing.

\begin{figure}
\includegraphics[width=9.1cm]{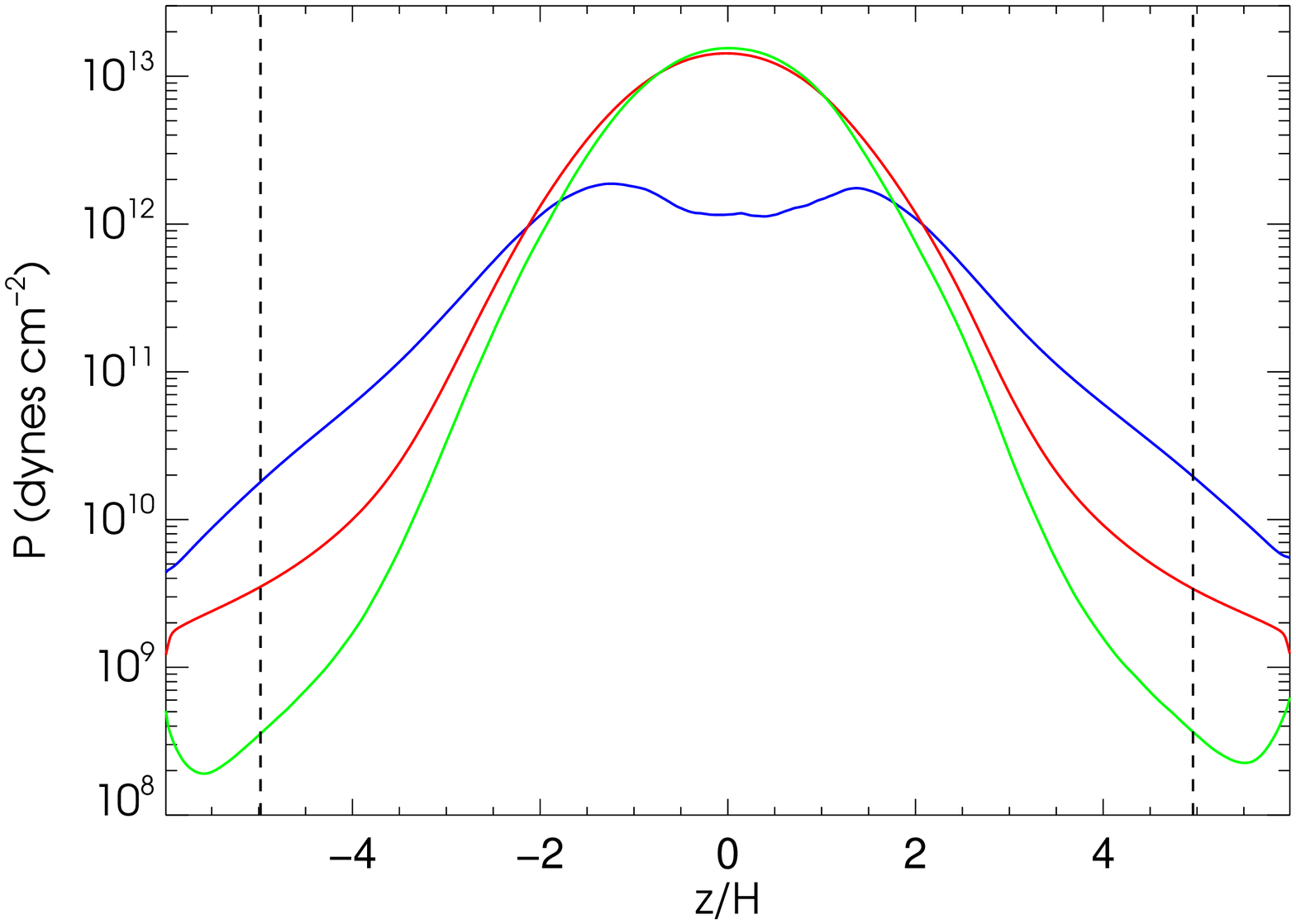}
\includegraphics[width=9cm]{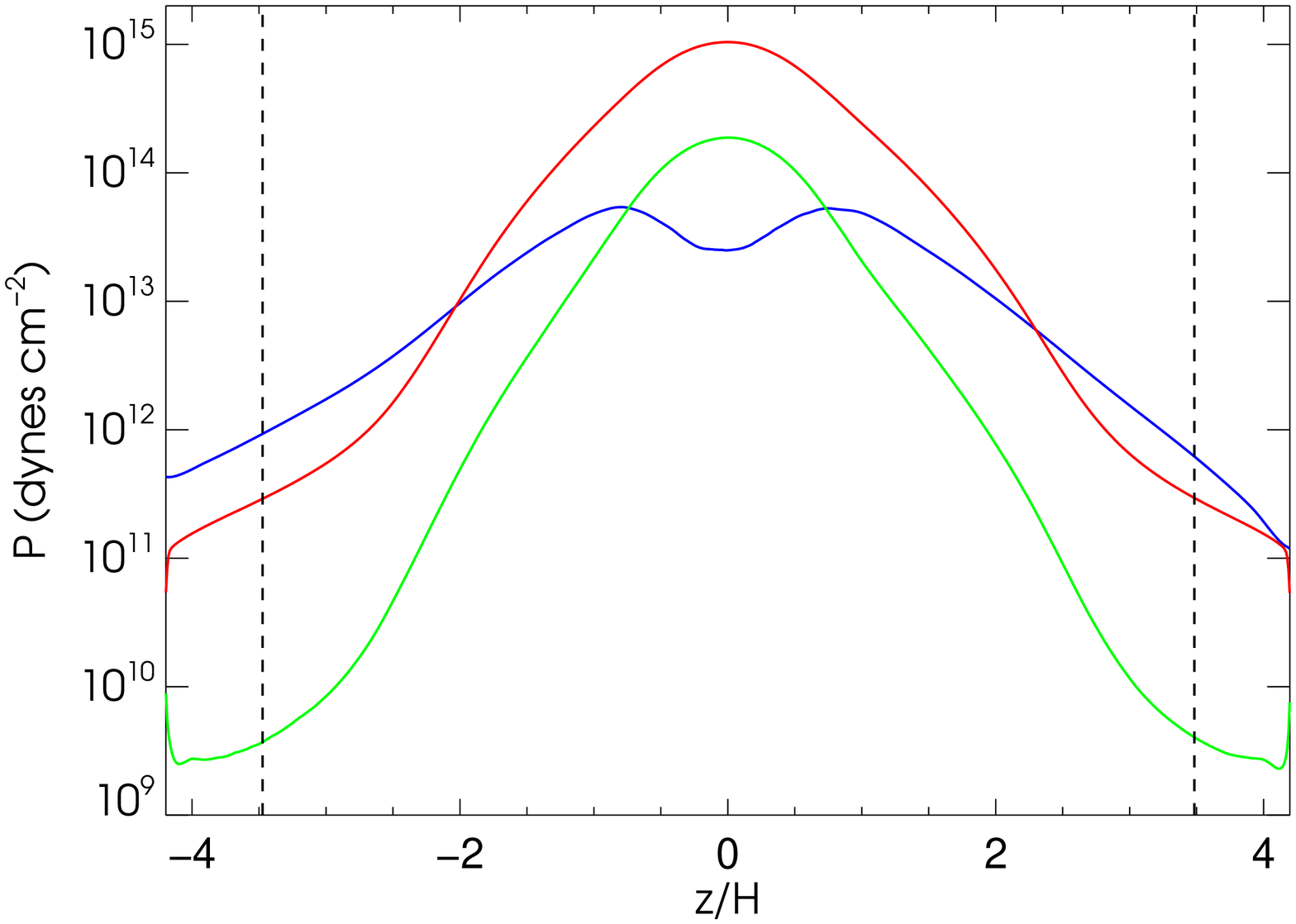}
\includegraphics[width=9cm]{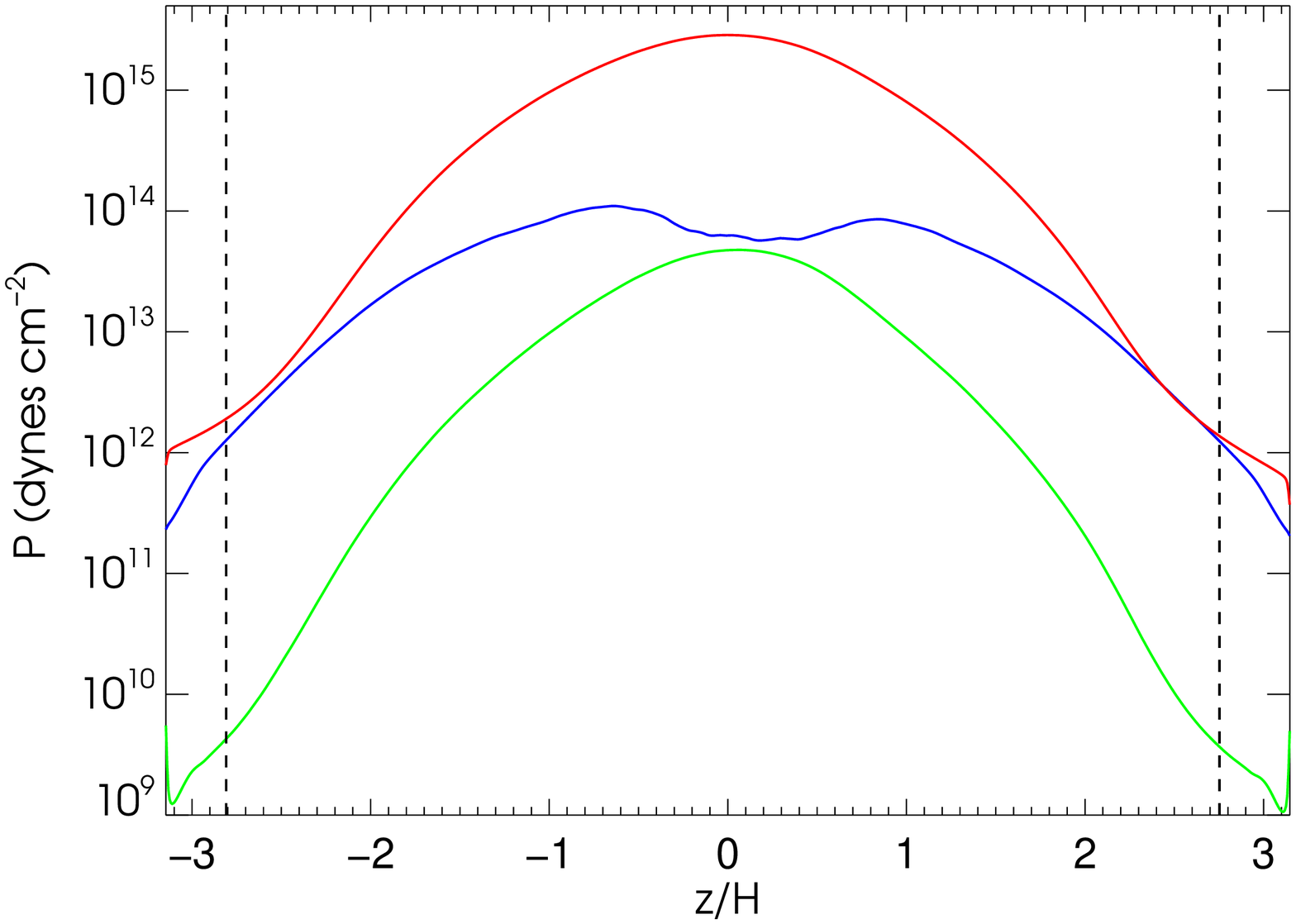}
\caption{Horizontal and time averaged radiation (red), magnetic (blue) and
gas (green) pressures as a function of height in local shearing box accretion
disk simulations 0528a (top left), 1112a (top right) and 0519b (bottom).
Vertical dashed lines indicate the locations of the Rosseland mean
(mostly electron scattering) photospheres in the time-averaged structures.
Note that the
radiation to magnetic pressure ratio in the disk surface layers increases with
increasing overall (volume integrated and time averaged) radiation to gas
pressure ratio.}
\label{fig:pressures}
\end{figure}

Figure~\ref{fig:pressures} depicts the time and horizontally-averaged vertical
profiles of radiation, gas and magnetic pressures in each of the three
simulations.  (The height $z$ on the horizontal axis in
these and other figures in this paper is in units of the scale height $H$
of the initial condition
of the simulations: approximately $3.1\times10^6$~cm for 0528a,
$1.46\times10^6$~cm for 1112a and $4.37\times10^6$~cm for 0519b.
This initial scale height is comparable to the actual total pressure scale
height of the resulting time-averaged structure.)
In the outer layers of the accretion disk in all three
simulations, magnetic pressure greatly dominates gas pressure, and also
generally dominates radiation pressure.  As a result, the Parker instability
generally controls the dynamics of these layers on
the scales resolved by the simulations \citep{bla07}.
However, Figure \ref{fig:pressures} shows that at the highest levels of
radiation to gas pressure ratio, radiation pressure is becoming comparable
to magnetic pressure in supporting the disk outer layers.
We therefore expect the photon bubble instability to be more prominent in
higher radiation pressure systems as the transition wavenumber $k_{\rm tran}$
decreases with decreasing magnetic to radiation pressure ratio, assuming
comparable radiation and magnetic pressure scale heights.
Photon bubbles, rather than Parker instability, may control 
the surface layer dynamics at high levels of radiation pressure support.

\begin{figure}
\includegraphics[width=8.5cm]{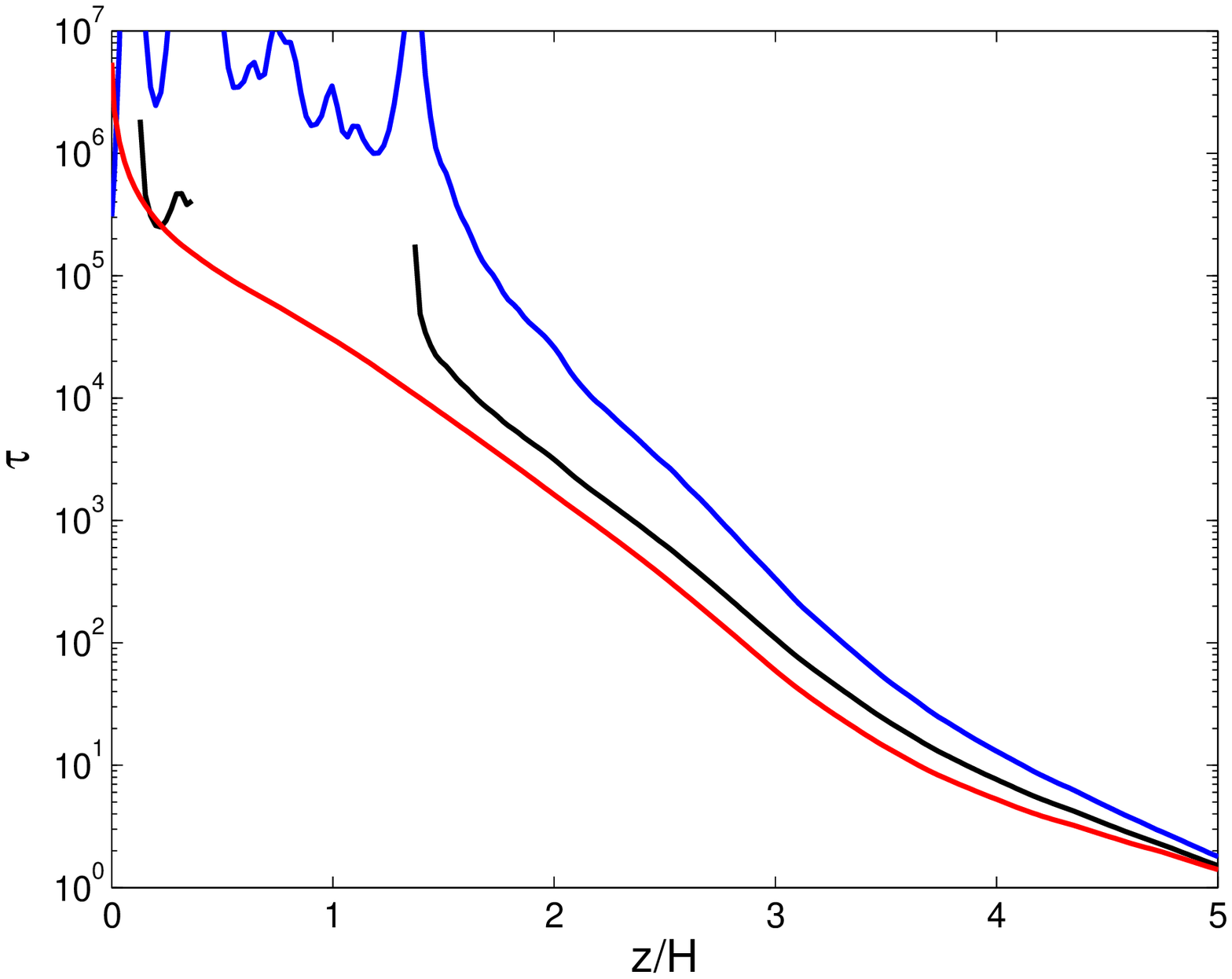}
\includegraphics[width=8.5cm]{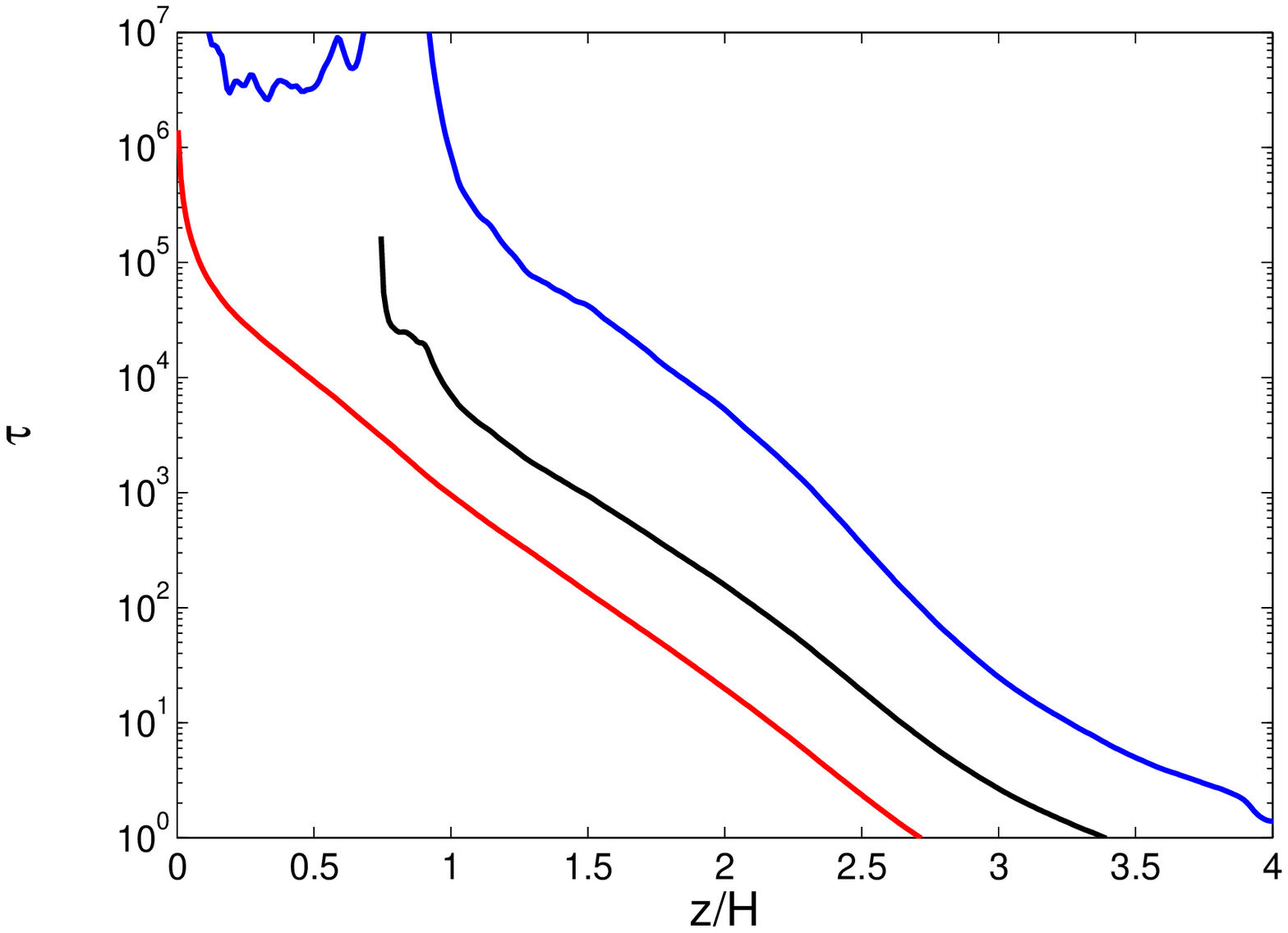}
\includegraphics[width=8.5cm]{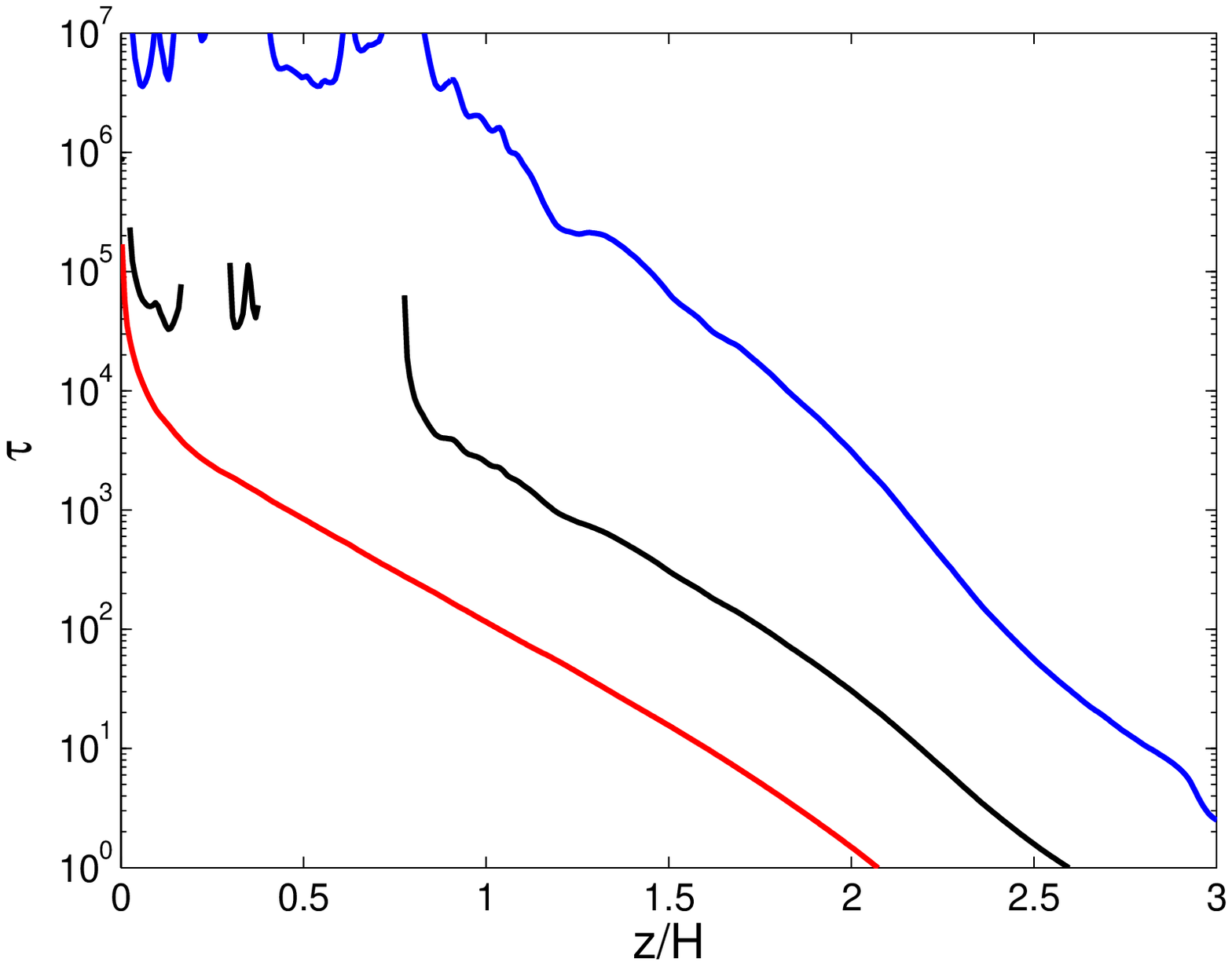}
\caption{Optical depth of the characteristic wavelengths to electron scattering
as a function of height for the horizontally and time-averaged structures in
simulations 0528a (top left), 1112a (top right),
and 0519b (bottom).  Red, black and blue curves represent the
photon bubble turnover wavelength, Parker cutoff wavelength and photon
bubble/Parker transition wavelength, respectively.}
\label{fig:tau}
\end{figure}

Beyond pressure considerations, the medium also needs to be optically thick
(but not so thick as to impede radiative diffusion) at a particular wavelength
in order to drive the photon bubble instability at that wavelength.
This is because diffusive radiative transport is essential for producing
radiative amplification when the opacity is dominated by Thomson scattering
\citep{bla03}.  However, Figure \ref{fig:tau} shows that the photon bubble
turnover wavelength becomes optically thin to electron scattering beyond
$|z/H|$ of approximately 2.1 and 2.7 for simulations 0519b and 1112a,
respectively. This
means that the instability cannot exist at wavenumbers higher than $k_{\rm T}$
and thus will not achieve its maximum growth rate near the disk photosphere.
These simple estimates mean that only longer wavelength photon bubble modes
can grow in the upper layers of such accretion disks.
Nevertheless, these slower, longer wavelength modes might still develop into
shock trains: approximate nonlinear solutions with finite optical depths
have been explored by \citet{beg06b}.  The Parker instability, on the
other hand, is not driven by radiative diffusion and hence is not subject to
such optical depth restrictions.

Incidentally, Figure~\ref{fig:tau} also shows that
$k_{\rm T}>k_{\rm P}>k_{\rm tran}$ in all three simulations.  With the
exception of 0528a (see Figure \ref{fig:0528ak}) where the gas pressure in
the disk upper layer is not negligibly small, the growth rate dependence on
wavenumber therefore resembles that depicted in
Figure~\ref{fig:finitegas}.  As shown in that figure, the
Parker instability transitions into the photon bubble instability at
short wavelengths before magnetic tension would introduce a cutoff.  Note,
however, that the photon bubble instability would never reach the asymptotic
growth rate depicted in that figure because the turnover wavelength
$2\pi/k_{\rm T}$ is optically thin. 

\begin{figure}
\includegraphics[width=9.5cm]{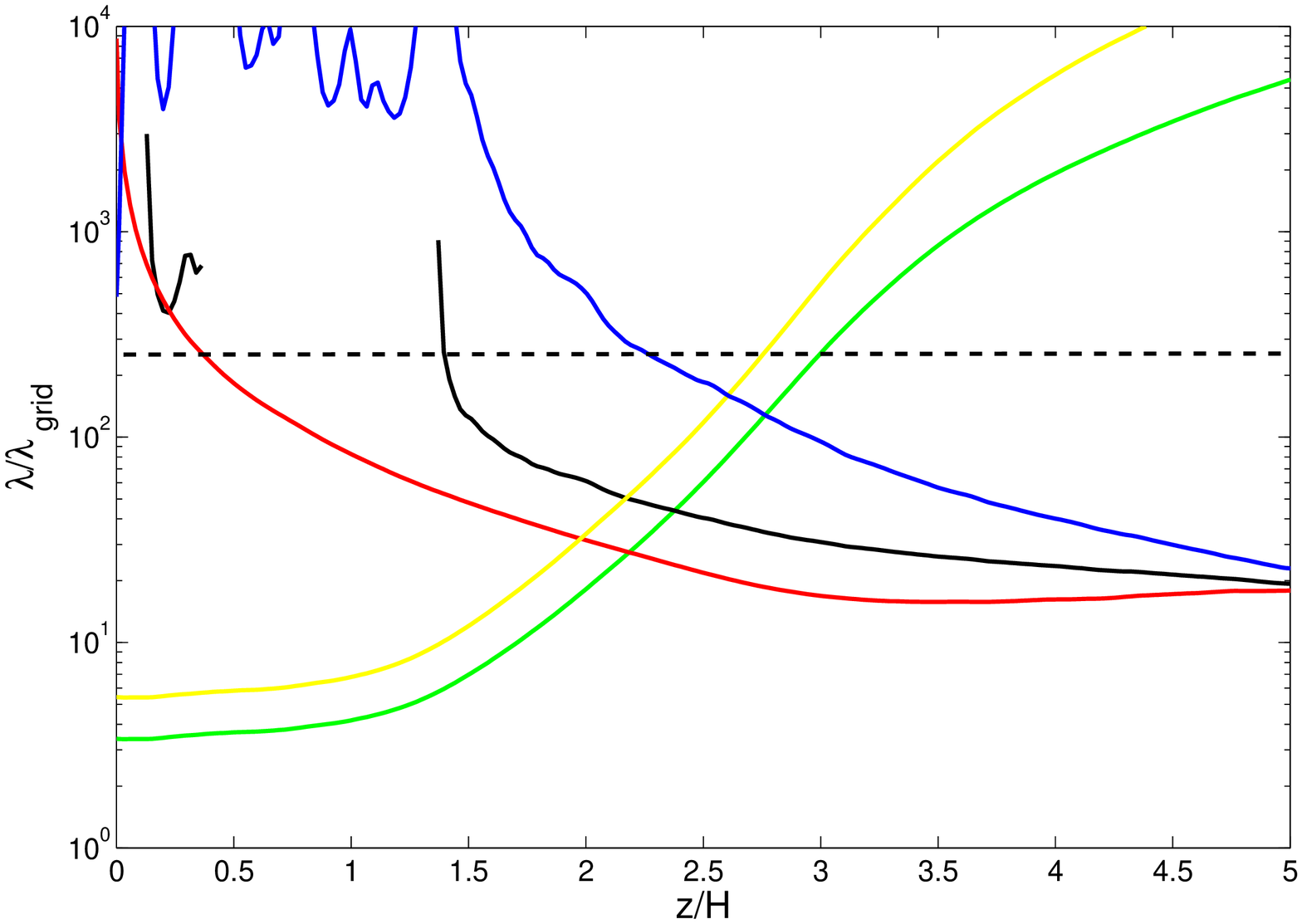}
\includegraphics[width=9.5cm]{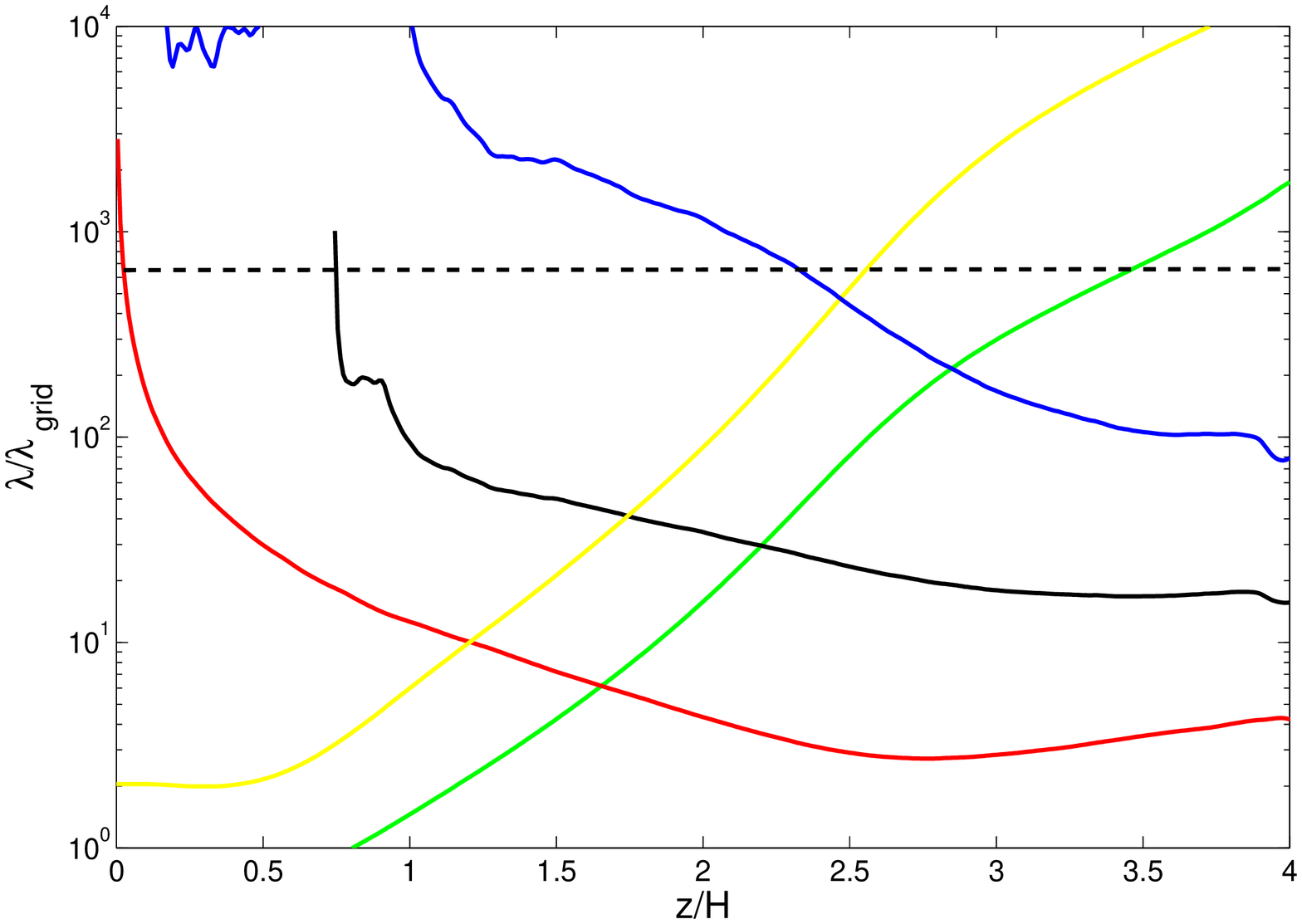}
\includegraphics[width=9.5cm]{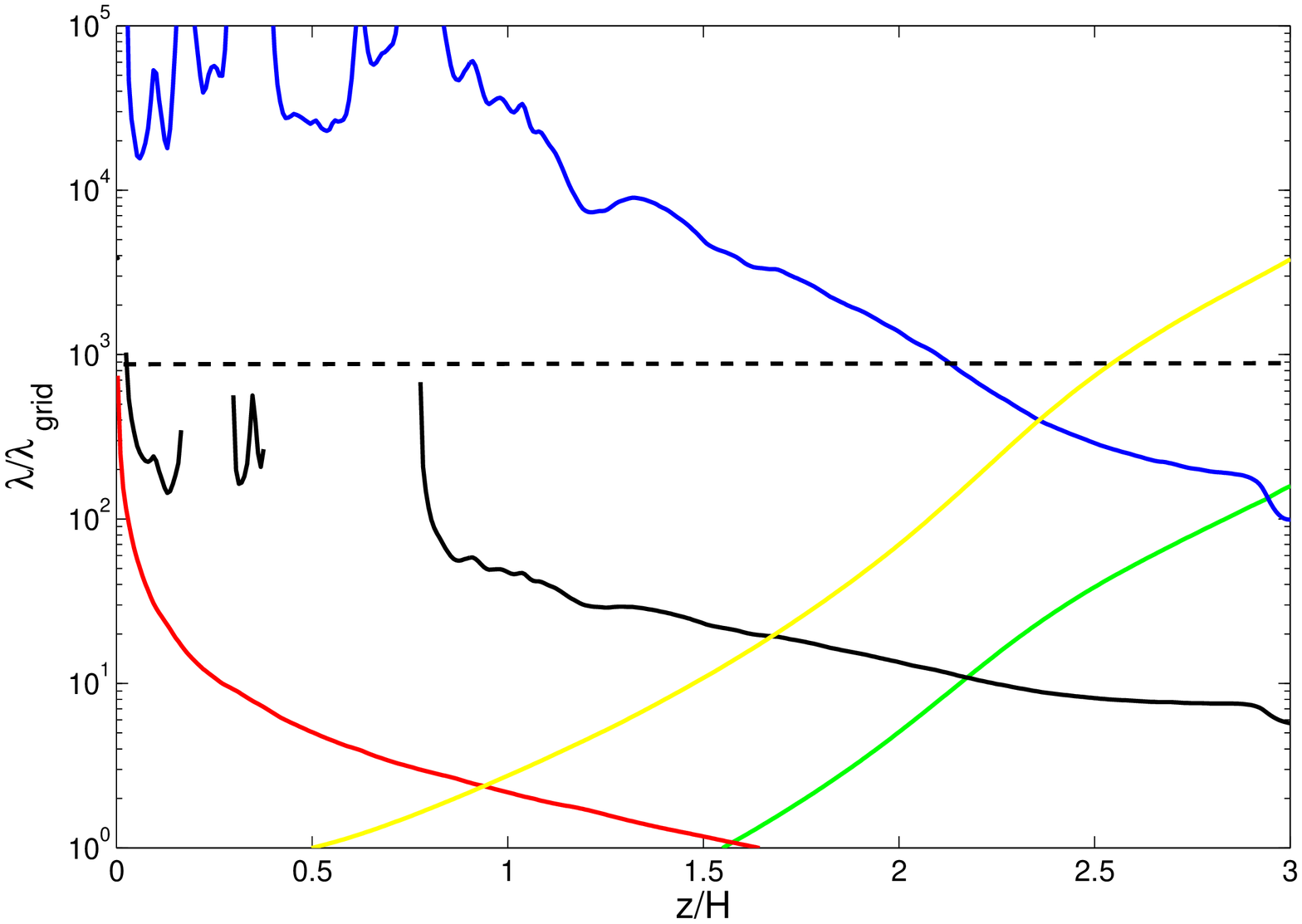}
\caption{The critical wavelengths $\lambda_{\rm tran}$ (blue),
$\lambda_{\rm P}$ (black) and $\lambda_{\rm T}$ (red), normalized by the
azimuthal grid cell size, as a function of height for the horizontally and
time-averaged structures of simulations 0528a (top left,
$P_{\rm rad}\cong P_{\rm gas}$), 1112a (top right,
$P_{\rm rad}\cong 7P_{\rm gas}$) and 0519b (bottom,
$P_{\rm rad}\cong 70P_{\rm gas}$). The green curves indicate the
wavelength $\lambda_{\rm R}$ below which radiative diffusion is rapid for
acoustic perturbations, and the yellow curves indicate the wavelength
$\lambda_{\rm S}$ above which radiative diffusion is slow for acoustic
perturbations.  The gaps in the black curves indicate
regions where the horizontal and time averaged magnetic pressure
increases vertically outward.  Our WKB analysis is only valid below the
horizontal dashed lines, which indicate the value of $2\pi H$ in each
of the simulations.}
\label{fig:lcrit}
\end{figure}

Figure~\ref{fig:lcrit} shows the same three characteristic wavelengths as
a function of height in the simulations, now scaled by the azimuthal grid
cell size.
Also shown are the characteristic wavelengths that demarcate the
rapid ($\lambda<\lambda_{\rm R}$) and slow ($\lambda>\lambda_{\rm S}$)
radiative diffusion regimes for acoustic perturbations \citep{bla07}:
\begin{equation}
\lambda_{\rm R}\equiv\frac{2\pi c}{3c_{\rm i}\kappa\rho}
\left(\frac{4E}{e+4E}\right)\left(\frac{c_{\rm i}}{c_{\rm t}
}\right)^2
\label{lr}
\end{equation}
and
\begin{equation}
\lambda_{\rm S}\equiv\frac{c_{\rm t}}{c_{\rm i}}\lambda_{\rm R}.
\label{ls}
\end{equation}
Most of the results of this paper only apply in the rapid diffusion regime
($\lambda<\lambda_{\rm R}$, i.e. below the green curves in
Figure~\ref{fig:lcrit}.  We therefore see that Parker instabilities with small
$k_x$ cannot exist in the rapid diffusion regime in the highest radiation
to gas pressure ratio simulation 0519b, as they have already transitioned into
photon bubbles.  This reflects the fact that we noted above that magnetic
pressure is not dominant over radiation pressure in the surface layers of
this simulation.  However, large $k_x$ modes (i.e. those with large radial
shear between the field lines) will still be Parker-like in character even in
the rapid radiative diffusion regime, and the magnetic field and density
structure in this simulation still shows evidence of Parker instability
\citep{bla11}.

Photon bubbles can reach their maximum growth rate in a
narrow range of depths in simulations 0519b and 1112a, where radiative damping
becomes small compared to the photon bubble asymptotic growth rate and the
turnover wavelength is optically
thick to scattering. Here the photon bubbles may evolve into nonlinear shock
trains \citep{beg01} that can propagate outward into the photosphere and
potentially produce observable signatures. However, Figure~\ref{fig:lcrit}
shows that the turnover wavelength in 1112a and 0519b is at most a few times
larger than the grid zone size in the region where $k_{\rm T}$ is in the rapid
diffusion limit, which means that the simulation cannot resolve the fastest
growing photon bubble modes, even if they can physically exist in this region.

\begin{figure}
%\plottwo{1112aGrowth.eps}{0519bGrowth.eps}
\plottwo{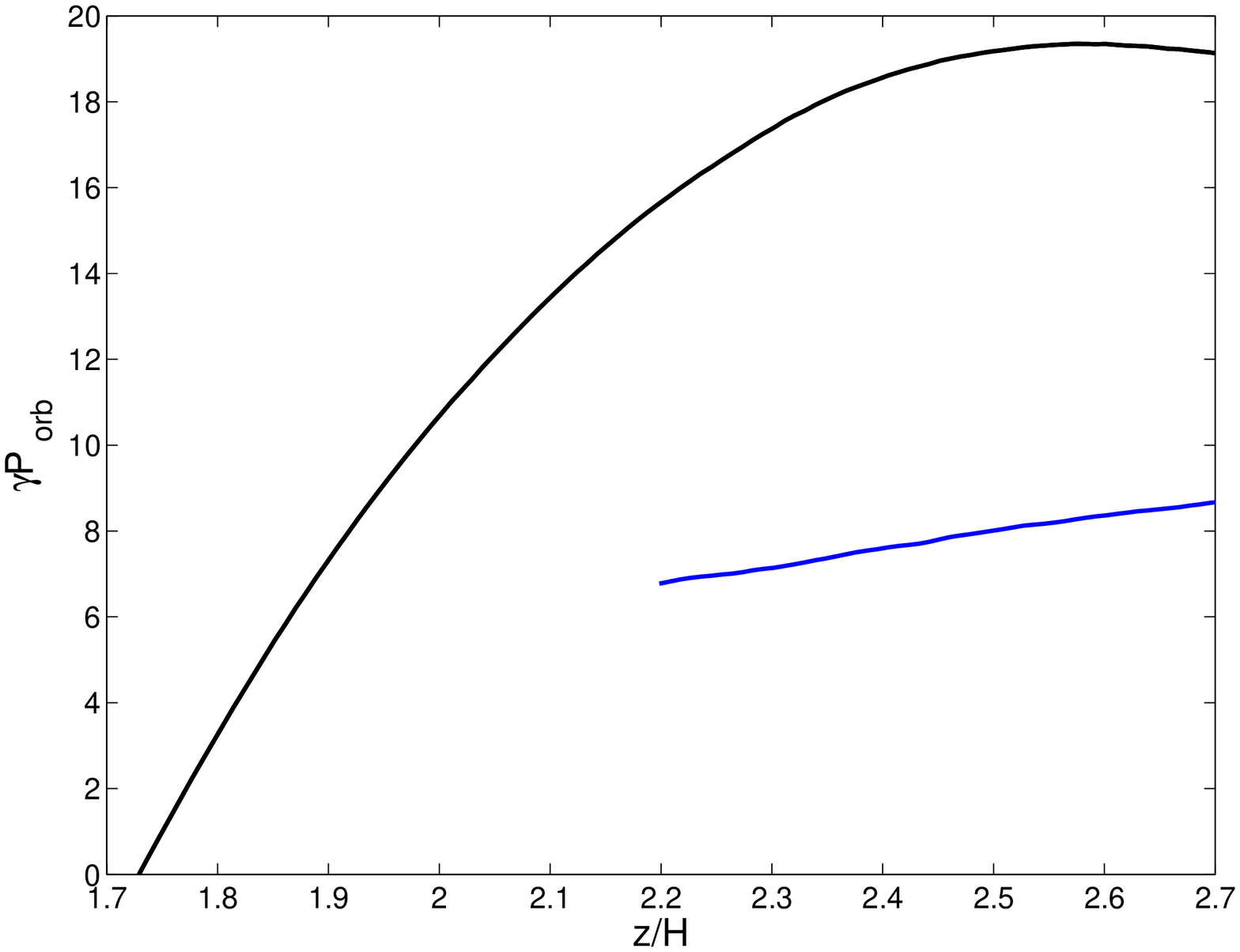}{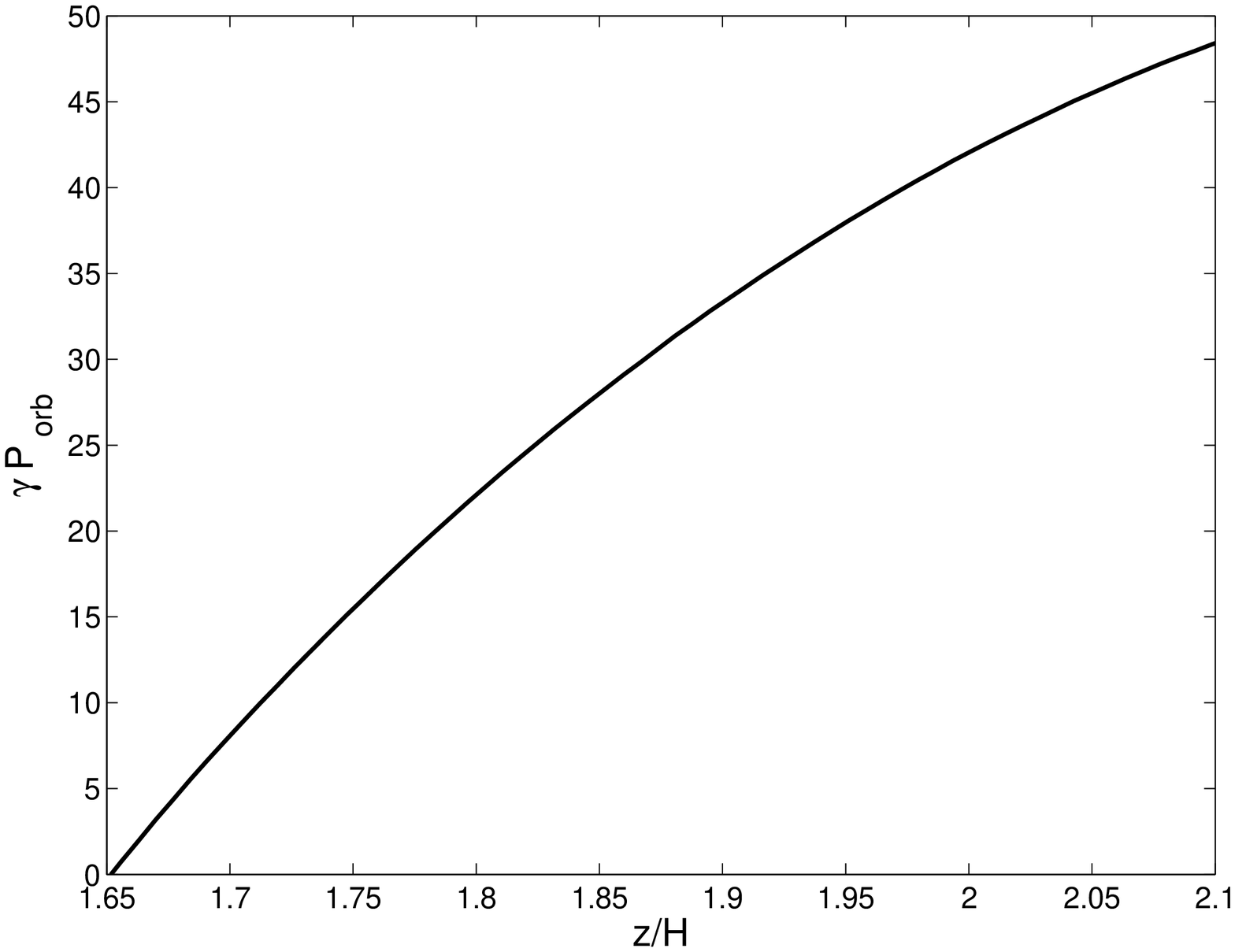}
\caption{Maximum rapid diffusion Parker growth rate (blue) [from equation
(\ref{eqgammamaxgilman})] and the asymptotic short-wavelength photon bubble
growth rate [black, based on equation (93) of \citet{bla03}] as a
function of height from horizontal and time averaged data above the midplane
in simulations 1112a (left) and 0519b (right).  The plotted growth rates are
scaled with the local orbital period for each simulation.  The wave vector
orientation is $\hat{\bf k}=(0,\cos(\pi/4),\sin(\pi/4))$. At heights below
those shown in the plots, photon bubble growth is suppressed by radiative
damping, while at heights above those shown, the photon bubble
turnover wavelength is optically thin, so that the short wavelength asymptotic
growth rates cannot be achieved.
Note that the plotted Parker growth rate is only valid within regions where
the cutoff wavelength $k_{\rm P}$ is in the rapid diffusion regime, which in
the case of 0519b is above the plotted range.}
\label{fig:highk}
\end{figure}

Figure \ref{fig:highk} shows the asymptotic, short wavelength, photon bubble growth rate (black curve), as well as the rapid diffusion Parker growth rate
(blue curve), in the radiation dominated simulations 1112a and 0519b.  The
plots are restricted to a range of heights where radiative damping of
photon bubbles is small enough so that they are unstable, and where the
turnover wavelength is optically thick.  Within this range, we only show
the Parker growth rates where the cutoff wavelength $\lambda_{\rm P}$ is in the
rapid diffusion regime, where the analysis of this paper is valid.  Photon
bubbles grow faster than Parker in this range of heights, and so in principle
could be important, but as we noted above, the simulations cannot resolve these
fastest growing wavelengths. Note that the midplane regions, where most of the
magnetorotational turbulence is acting, are not photon bubble unstable.

\begin{figure}
%\plotone{0528aGrowth.eps}
\plotone{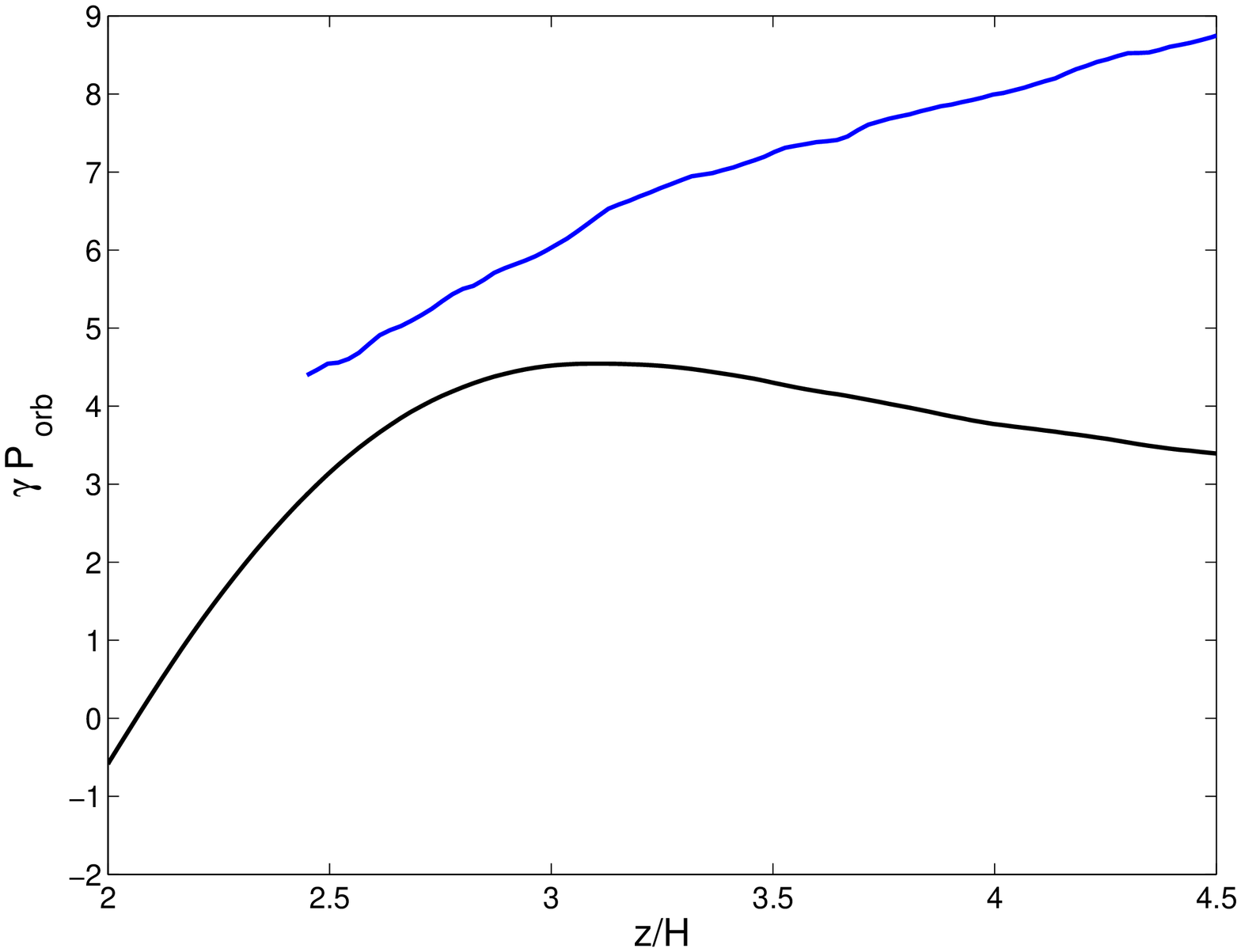}
\caption{Same as Figure 7 but for simulation 0528a.}
\label{fig:0528agrowth}
\end{figure}

Figure~\ref{fig:0528agrowth} shows the same information for simulation 0528a,
and may explain why \cite{bla07} did not observe the photon bubble instability
in this $P_{\rm rad}\approx P_{\rm gas}$ simulation. There the transition
wavelength in the upper layers is moderately optically thick to electron
scattering, which suggests that photon bubbles can physically exist in the
simulated disk.  However, Figure~\ref{fig:0528agrowth} shows that Parker is the
dominant instability, at least for $|z/H|\gtrsim2.5$, where $\lambda_{\rm P}$ is
in the rapid diffusion regime. Figure \ref{fig:0528ak} shows the growth rate as
a function of wavenumber at a representative height ($z/H=3.5$) in the
disk upper layer and further illustrates this conclusion. Closer to the midplane,
radiative damping dominates over photon bubbles in the short wavelength limit.
Note that our plotted results are comparable to and consistent with the growth
rates estimated by \cite{bla07} at two specific epochs in this simulation.
Those authors did not know how Parker and photon bubbles couple together, but
we have demonstrated here that the maximum growth rates of the individual
instabilities are in fact unaltered.

\begin{figure}
%\plotone{0528aZ3.5H.eps}
\plotone{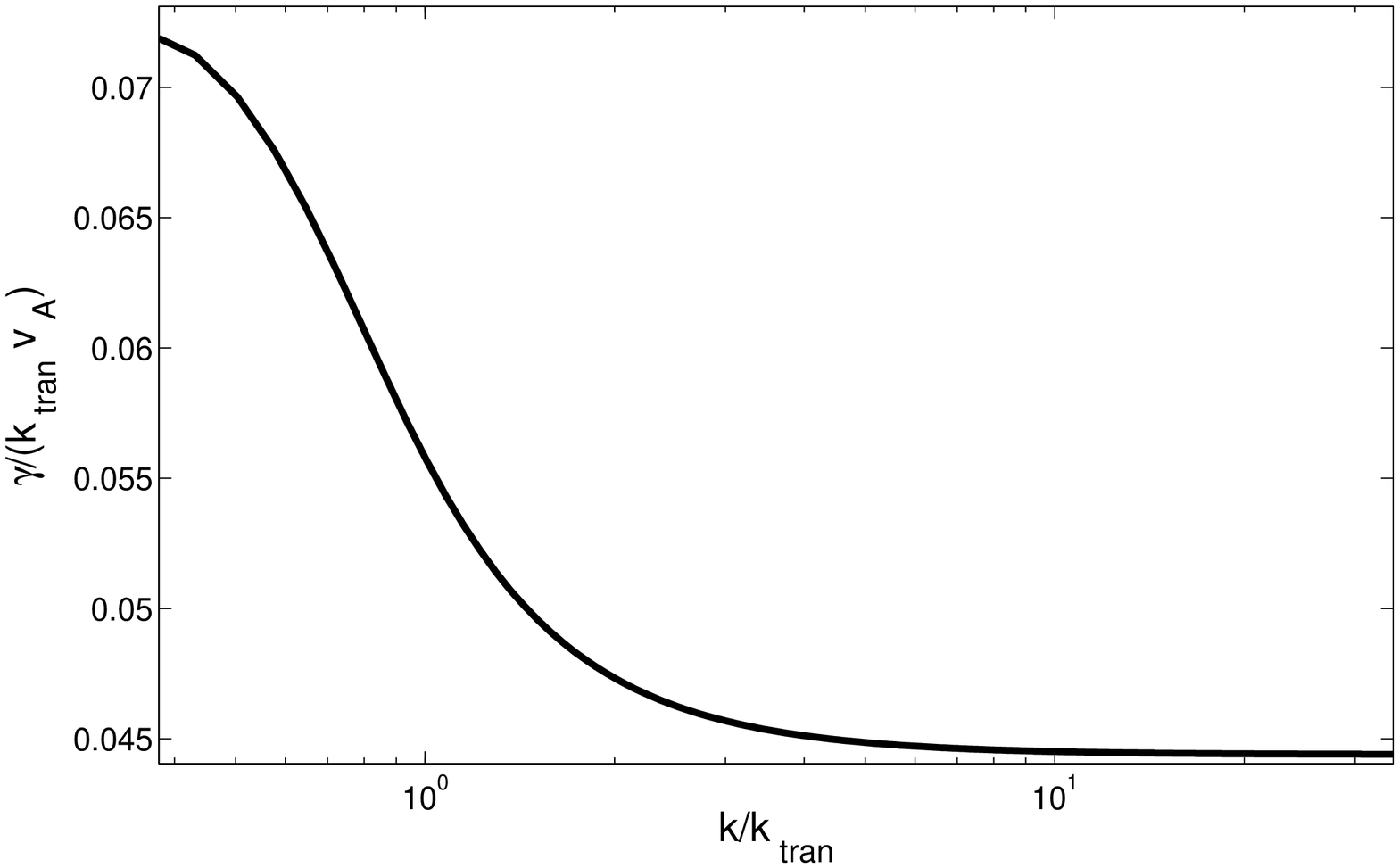}
\caption{Instability growth rate as a function of wavenumber for parameters
taken from the horizontal and time averaged simulation 0528a results at
$z/H=3.5$. We see that the maximum Parker instability growth rate is indeed
higher than the photon bubble asymptotic growth rate in this case, in
agreement with Figure \ref{fig:0528agrowth}.}
\label{fig:0528ak}
\end{figure}

\section{Conclusions}

We expanded previous studies of photon bubble instabilities to include
background magnetic field gradients, thereby introducing a coupling between
photon bubble and Parker instabilities.
Our first main result was obtained assuming negligible gas pressure in
the equilibrium, which in principle allows the photon bubble and Parker
instabilities to exist out to very short wavelengths.  There we identified a
finite transition wavenumber
$k_{\rm tran}$ between the photon bubble and Parker instabilities. The
instability is Parker-like for $k<k_{\rm tran}$ and photon bubble-like for
$k>k_{\rm tran}$, although strong horizontal shear between neighboring magnetic
field lines (i.e. large $k_x$) pushes the Parker-like range to higher
wavenumber.

We then proceeded to a naive WKB study including finite gas pressure. We
numerically found approximately
the same transition wavenumber. Moreover, finite gas pressure introduces
finite photon bubble turnover and Parker cutoff wavenumbers ($k_{\rm T}$ and
$k_{\rm P}$, respectively).  For $k_{\rm P}<k_{\rm tran}$, we see a sharp
growth rate decrease above $k\simeq k_{\rm P}$, while the photon bubble
growth rate becomes constant in $k$ for
$k>k_{\rm T}$. Once again, the growth rates of both instabilities agree with
previous results in all wavenumber regimes. We also derived the scaling of the
important wavenumbers $k_{\rm T}$, $k_{\rm P}$ and $k_{\rm tran}$ as functions
of background parameters in equations (\ref{ktran})-(\ref{kp}).
Note that the results of this paper can only be applied to backgrounds where
the characteristic wavenumbers $k_{\rm tran}$, $k_{\rm P}$ and $k_{\rm T}$
satisfy the WKB condition $kH\gg 1$.

We also developed a WKB analysis of the coupled Parker/photon bubble problem
including the effects of differential rotation.  Photon bubbles are not
significantly affected by shear and rotation provided their wavelengths are
short enough.  On the other hand, Parker modes can be affected by shear
unless they have short radial wavelengths.  This is equivalent to the
condition of strong horizontal shear between neighboring field lines that
expands the Parker-like range to higher wavenumber.

We applied our results to analyze radiation MHD shearing box accretion disk
simulations 0528a, 1112a and 0519b with volume integrated radiation to gas
pressure ratios of 1, 7 and 70, respectively. We found that the
asymptotically growing short wavelength photon bubble instability is not
likely to exist in the upper layers of both 1112a and 0519b because the
turnover wavelength becomes optically thin to electron scattering in the
photosphere. For 0528a, the undulatory Parker instability in the rapid
diffusion regime dominates over photon bubbles in the upper layers. On the
other hand, radiative damping destroys the photon bubbles near the midplane
in all three simulations, and photon bubbles may exist at the asymptotic
growth rate only in a narrow range of heights.  Even there it is not clear
whether the turbulent conditions would allow the photon bubbles to grow
unimpeded. Last but not least, the grid cell sizes of current radiation
pressure dominated simulations are inadequate to fully resolve the turnover
wavelength of the photon bubble instability,
so that the fastest growing wavelengths cannot manifest themselves. Higher
resolution radiation pressure dominated simulations may reveal this
instability.

While accretion disk applications have been our focus, much of the analysis
of this paper applies to static media, and magnetized stellar envelopes
are clearly another area where this physics would be relevant.  Generalizing the
simulations by \citet{tur05} of magnetized, vertically stratified static
equilibria to include magnetic pressure gradients would be interesting
in order to explore how the photon bubble and Parker instabilities interact
in the nonlinear regime.  Even if short wavelength photon bubbles have faster
linear growth rates than the longer wavelength Parker modes, \citet{tur05} have
shown that the resulting nonlinear shock trains merge, causing the distances
between adjacent shocks and the density contrast to grow until the
magnetic field buckles (or until only one wavelength fills the simulation
domain).  This field line buckling may resemble the nonlinear development
of the undulatory Parker instability, and it may be that both instabilities
ultimately lead to a similar nonlinear outcome, at least in some regimes.

Finally, we stress an important caveat to our work:  the photon bubble-Parker
transition regime of our numerical finite gas pressure results do not readily
apply when $c_{\rm i}\geq v_{\rm A}$, $c_{\rm r}$ or $c_{\rm r}\geq v_{\rm A}$,
due to limitations of the WKB method discussed in section 4. Such regimes can
exist in the portion of the accretion disk within and just outside the MRI
unstable region, where the gravitational
potential energy is dissipated into heat \citep{hir09}. The behavior of the
Parker-photon bubble transition in this region may be important for
understanding how accretion power is transported outward.  This is a difficult
problem that may deserve further future work.

We owe a debt of gratitude to Ellen Zweibel for illuminating many aspects of
the physics of the Parker instability.  We also thank the referee for
constructive criticism and for suggesting that we look more carefully at
the issues of rotation and shear when considering accretion disk applications.
We have benefited from helpful discussions
with L. Bildsten, M. Block, K. Choiu, E. Gallo, J. Jacob, E. Rykoff, N. Turner,
E. Newton and D. Harsono. This work was supported in part by NSF grants
AST-0307657 and AST-0707624.

\appendix 

\section{Magnetic Buoyancy Instabilities in the Absence of Photon Bubbles}
\label{parkerappendix}

In order to better understand the coupled photon bubble-Parker problem, it is
useful to derive the properties of pure Parker modes in the absence
of photon bubble instabilities.  We are most interested in a medium where radiation pressure dominates gas pressure, and in short wavelengths
where radiative diffusion is rapid.  The latter condition implies that
temperature perturbations are very small, and if we take them literally to be
zero, then the flux perturbations that drive photon bubble instabilities in
this regime will vanish identically, leaving only pure Parker modes.

Parker modes with rapid radiative diffusion were first studied by \citet{gil70},
who argued that it was a good first order approximation to replace the
perturbed energy equation (\ref{eqenergypert}) with the statement that the
temperature perturbation $\delta T$ vanishes.

The linearized perturbation equations that we presented in subsection 2.2 can
be combined with the perturbed flux-freezing equation to give the total (gas
plus radiation plus magnetic) pressure perturbation in two radiative diffusion
limits. The first is the adiabatic limit of infinitely slow diffusion
($\kappa\rightarrow\infty$, $F\rightarrow0$, $\kappa F$ finite), giving
\begin{eqnarray}
\delta P_{\rm tot}&=&\rho\left(c_{\rm t}^2+v_{\rm A}^2-{k_y^2v_{\rm A}^2\over
\omega^2}c_{\rm t}^2\right)\tilde{\delta\rho}-\rho\left[
{c_{\rm t}^2N^2\over g}+v_{\rm A}^2{d\over dz}\ln\left({B\over\rho}\right)
\right]\xi_z\cr
&+&\rho\left({k_y^2v_{\rm A}^2\over\omega^2}\right)
\left[{c_{\rm t}^2N^2\over g}+v_{\rm A}^2{d\over dz}\ln B\right]\xi_z,
\label{eqdptotadiabatic}
\end{eqnarray}
where $\xi_z=\int\delta v_z dt=i\delta v_z/\omega$ is the vertical component of
the Lagrangian displacement vector. The second is the limit of sufficiently
short wavelengths that diffusion is rapid enough ($\kappa\rightarrow0$) to
guarantee that the perturbations are isothermal ($\delta T\rightarrow0$). This
gives
\begin{eqnarray}
\delta P_{\rm tot}&=&\rho\left(c_{\rm i}^2+v_{\rm A}^2-{k_y^2v_{\rm A}^2\over
\omega^2}c_{\rm i}^2\right)\tilde{\delta\rho}-\rho
v_{\rm A}^2{d\over dz}\ln\left({B\over\rho}\right)\xi_z\cr
&+&\rho\left({k_y^2v_{\rm A}^2\over\omega^2}\right)
\left[v_{\rm A}^2{d\over dz}\ln B\right]\xi_z.
\label{eqdptotrapiddiff}
\end{eqnarray}

Equations (\ref{eqdptotadiabatic}) and (\ref{eqdptotrapiddiff}) contain
the essential physics of the Parker instability in both of these limits.
Buoyancy is maximized for a vertically displaced fluid element if it is
in pressure equilibrium with its surroundings, i.e. $\delta P_{\rm tot}=0$.
Solving the resulting equation for the Eulerian density perturbation
associated with an upward fluid element displacement ($\xi_z>0$) then
gives instability if $\tilde{\delta\rho}<0$.  There are two possibilities
depending on $k_y$.  For $k_y\rightarrow0$, the first $\xi_z$ term
on the right hand sides of equations (\ref{eqdptotadiabatic}) and
(\ref{eqdptotrapiddiff}) dominates.  In this case, the perturbation involves
a straight bundle of field lines, and the resulting mode is called an
interchange mode.  On the other hand, for $k_y$ sufficiently large, 
the second $\xi_z$ term dominates, and the perturbation involves a bending
of the field lines, resulting in the undulatory mode.

In the adiabatic limit, the instability criteria for these two modes
\citep{new61,ach79,chr03} are therefore
\begin{eqnarray}
{-g\over c_{\rm t}^2}{d\over dz}\ln\left({B\over\rho}\right)&>&
{N^2\over v_{\rm A}^2}\,\,\,\,\,{\rm interchange}\cr
{-g\over c_{\rm t}^2}{d\over dz}\ln B&>& {N^2\over v_{\rm A}^2}
\,\,\,\,\,{\rm undulatory}.
\label{newcombcriteria}
\end{eqnarray}

In the rapid diffusion limit, on the other hand, the instability criteria
may be written as \citep{gil70,ach79}
\begin{eqnarray}
{-g\over c_{\rm i}^2}{d\over dz}\ln\left({B\over\rho}\right)&>&
0\,\,\,\,\,{\rm interchange}\cr
{-g\over c_{\rm i}^2}{d\over dz}\ln B&>& 0
\,\,\,\,\,{\rm undulatory}.
\label{gilmancriteria}
\end{eqnarray}
These are easier to satisfy than the criteria (\ref{newcombcriteria})
because pressure and temperature equilibrium between the perturbations and
their surroundings makes the fluid hydrodynamically neutrally buoyant.

We consider only the rapid diffusion limit from now on.  Using
equation (\ref{eqdptotrapiddiff}), the linearized
continuity and horizontal momentum equations can be combined to express
the density and total pressure perturbations entirely in terms
of the vertical velocity perturbation:
\begin{equation}
\tilde{\delta\rho}={-i\omega\over k_{\perp}^2c_{\rm i}^2-\omega^2
{\omega^2-k_{\perp}^2v_{\rm A}^2\over\omega^2-k_y^2v_{\rm A}^2}}
\left[{\omega^2-k_{\perp}^2v_{\rm A}^2\over\omega^2-k_y^2v_{\rm A}^2}
{\delta v_z\over H_\rho}+{k_\perp^2v_{\rm A}^2\over2\omega^2H_{\rm mag}}
\delta v_z-{d\delta v_z\over dz}\right]
\label{eqdrhorapiddiff}
\end{equation}
and
\begin{equation}
\delta P_{\rm tot}={-i\omega\rho\over k_{\perp}^2c_{\rm i}^2-\omega^2
{\omega^2-k_{\perp}^2v_{\rm A}^2\over\omega^2-k_y^2v_{\rm A}^2}}
\left[{c_{\rm i}^2\over H_\rho}\delta v_z+{v_{\rm A}^2\over2H_{\rm mag}}
\delta v_z-\left(c_{\rm i}^2+v_{\rm A}^2-{k_y^2v_{\rm A}^2\over\omega^2}
c_{\rm i}^2\right){d\delta v_z\over dz}\right].
\label{eqdptotrapiddiff1}
\end{equation}
Here $k_\perp\equiv(k_x^2+k_y^2)^{1/2}$ is the magnitude of the horizontal
wavenumber.

These expressions can then be combined with the linearized vertical momentum
equation,
\begin{equation}
-i\omega\rho\delta v_z=-{\partial\delta P_{\rm tot}\over\partial z}
-\rho\tilde{\delta\rho}g-{ik_y^2\over\omega}\rho v_{\rm A}^2\delta v_z,
\end{equation}
to give an exact linear ordinary differential equation for the vertical
velocity perturbation.  In his analysis of the Parker instability in
the limit of rapid radiative diffusion, \citet{gil70} derived a simplified
form of this differential equation by first assuming $k_x\rightarrow\infty$
with $\omega$ and $k_y$ remaining finite.  From the $x$-component of the
linearized momentum equation,
\begin{equation}
-i\omega\rho\delta v_x=-ik_x\delta P_{\rm tot}
-{ik_y^2\over\omega}\rho v_{\rm A}^2\delta v_x,
\end{equation}
this guarantees that $\delta P_{\rm tot}\rightarrow0$, giving a
maximum buoyant response and therefore a maximum Parker growth rate.
However, here we will allow for the possibility that $k_y/k_x$ is not
necessarily small in magnitude as $k_x$ gets large.  This is
because the Parker cutoff wavenumber
$k_{\rm P}=[g/(2c_{\rm i}^2H_{\rm mag})]^{1/2}$ can be very
large in a radiation dominated plasma, in which the isothermal sound
speed in the gas alone $c_{\rm i}$ can be very small.  (In contrast,
in the adiabatic limit, the Parker cutoff wavenumber is comparable to
the inverse scale height of the background medium.)  Because a
plasma whose thermal pressure is dominated by radiation is necessarily
supported against gravity by radiation pressure and/or magnetic pressure
gradients, the Parker cutoff wavenumber in the rapid diffusion limit can
be much larger than the inverse scale height of the background.  It
therefore makes sense to consider the possibility of very large, as well
as very small,  $k_y$.

We therefore take $|\omega^2|\ll k_\perp^2v_{\rm A}^2$, but not necessarily
much smaller than $k_y^2v_{\rm A}^2$.  The first
derivative term in equation (\ref{eqdrhorapiddiff}) is higher order
than the other terms for short wavelengths, so the density perturbation
becomes
\begin{equation}
\tilde{\delta\rho}\simeq{-i\omega v_{\rm A}^2\delta v_z\over\omega^2
(c_{\rm i}^2+v_{\rm A}^2)-k_y^2v_{\rm A}^2c_{\rm i}^2}\left(
{\omega^2-k_y^2v_{\rm A}^2\over2\omega^2H_{\rm mag}}
-{1\over H_\rho}\right).
\end{equation}
For the total pressure perturbation in equation (\ref{eqdptotrapiddiff1}),
the first derivative term dominates, and
\begin{equation}
\delta P_{\rm tot}\simeq {i\rho(\omega^2-k_y^2v_{\rm A}^2)\over\omega
k_\perp^2}{\partial \delta v_z\over\partial z}
\end{equation}

Substituting into the vertical momentum equation, employing the short
wavelength WKB approximation $\partial/\partial z\rightarrow ik_z$, and
simplifying, we finally obtain the dispersion relation
\begin{eqnarray}
(v_{\rm A}^2+c_{\rm i}^2)\omega^4&-&\left[k_y^2v_{\rm A}^2
(v_{\rm A}^2+2c_{\rm i}^2)+{gv_{\rm A}^2k_\perp^2\over k^2}
\left({1\over H_\rho}-{1\over2H_{\rm mag}}\right)\right]\omega^2\cr
&+&k_y^2v_{\rm A}^4\left(k_y^2
c_{\rm i}^2-{gk_\perp^2\over 2H_{\rm mag}k^2}\right)=0,
\label{eqngilman}
\end{eqnarray}
where $k\equiv(k_\perp^2+k_z^2)^{1/2}$ is the total wavenumber magnitude of
the perturbation.
Apart from the generalization that $k_y$ and $k_z$ no longer need be
considered small compared to $k_x$, this is identical to equation (14) of
\citet{gil70}.

If either of the instability criteria (\ref{gilmancriteria}) are satisfied,
then all nonzero wavenumbers $k_y$ less than
$k_\perp k_{\rm P}/k$ will be unstable, where, again, the
characteristic Parker cut off wavenumber is defined by
\begin{equation}
k_{\rm P}^2\equiv{g\over2c_{\rm i}^2H_{\rm mag}}.
\label{eqkp2gilman}
\end{equation}
The maximum undulatory instability growth rate is given by
\begin{equation}
\gamma_{\rm max}^2={k_\perp^2g\over k^2v_{\rm A}^2}\left[{c_{\rm i}\over
H_\rho^{1/2}}-\left({c_{\rm i}^2\over H_\rho}+{v_{\rm A}^2\over2H_{\rm mag}}
\right)^{1/2}\right]^2,
\label{eqgammamaxgilman}
\end{equation}
and occurs at a wavenumber $k_0$ given by
\begin{equation}
k_0^2={k_\perp^2g\over k^2v_{\rm A}^4}\left[\left({c_{\rm i}^2\over
H_\rho}+{v_{\rm A}^2\over2H_{\rm mag}}\right)^{1/2}-{c_{\rm i}\over
H_\rho^{1/2}}\right]\left[{v_{\rm A}^2+c_{\rm i}^2\over c_{\rm i}H_\rho^{1/2}}
-\left({c_{\rm i}^2\over H_\rho}+{v_{\rm A}^2\over2H_{\rm mag}}\right)^{1/2}
\right].
\label{eqk0}
\end{equation}

Assuming that the gas sound speed is much less than the Alfv\'en speed
(the regime of interest for this paper), equations (\ref{eqgammamaxgilman})
and (\ref{eqk0}) become
\begin{equation}
\gamma_{\rm max}^2={k_\perp^2g\over2\hmag k^2},
\label{eqgilmancizero}
\end{equation}
and
\begin{equation}
k_0^2={k_\perp^2g\over k^2v_{\rm A}c_{\rm i}(2H_{\rm mag}H_\rho)^{1/2}}.
\label{eqk0cizero}
\end{equation}

\section{The Zero Gas Pressure Limit with Rotation and Shear}
\label{secrotationderivation}

Here we generalize the zero gas pressure limit analysis of
section~\ref{sectionpzero} to include the effects of rotation and
shear on the coupled Parker/photon bubble instability problem in a
differentially rotating accretion disk.  Because we are interested
in short wavelength perturbations, we focus on a local comoving patch of
the accretion disk, and employ the shearing box approximation
\citep{gol65,haw95}.  The fluid equations (\ref{eqcont})-(\ref{eqstate})
remain the same, except for two modifications.

The first is the addition of the terms
$-2\rho\Omega\hat{\bf z}\times{\bf v}+2q\rho\Omega^2x\hat{\bf x}$
to the right hand side of the momentum equation~(\ref{eqmom}), which
represent the combined effects of Coriolis, centrifugal, and radial
gravitational forces on the flow.  Here $\Omega$ is the angular velocity
of the local patch of disk and $q$ is a shear parameter, representing 
a variation of angular velocity with radius $r$ of $\Omega\propto r^{-q}$
($q=3/2$ for Keplerian shear).  We employ a Cartesian basis
$(\hat{\bf x},\hat{\bf y},\hat{\bf z})$, with $\hat{\bf x}$ in the local
radial direction, $\hat{\bf y}$ in the local azimuthal direction, and
$\hat{\bf z}$ in the vertical direction just as in the static equilibrium
studied in the main body of the paper.  We continue to write the vertical
gravitational acceleration as ${\bf g}=-g\hat{\bf z}$, with $g=\Omega^2 z$,
the equilibrium radiation flux as ${\bf F}=F\hat{\bf z}$, and the equilibrium
magnetic field as ${\bf B}=B(z)\hat{\bf y}$, i.e. a purely azimuthal magnetic
field.  The equilibrium state continues to be described by equations
(\ref{eqhydrostatic})-(\ref{eqraddiff}), but in addition there is now
an azimuthal flow velocity of ${\bf v}=-q\Omega x\hat{\bf y}$.

The second modification concerns horizontal boundary conditions.  In the
static analysis, we considered a horizontally homogeneous equilibrium, and
perturbations that were periodic in the horizontal direction
($\propto\exp[i(k_x x+k_yy)]$).  The
shearing box has all fluid variables subject to shearing periodic
boundary conditions in the radial direction \citep{haw95}.  The standard
way to treat this is to make a coordinate transformation:  $x^\prime=x$,
$y^\prime=y+q\Omega xt$, $z^\prime=z$, and $t^\prime=t$.  Our perturbations
can then be taken to have a horizontal dependence of the form
$\exp[i(k_x(t^\prime)x+k_y y)]$, where the wavenumber in the $x$-direction
is time-dependent:
\begin{equation}
k_x(t^\prime)=k_{x0}+q\Omega t^{\prime}k_y
\end{equation}
Here $k_{x0}$ and $k_y$ are independent of time.
Horizontal derivatives acting on perturbations can therefore be written as
$\partial/\partial x\rightarrow ik_x(t^\prime)$ and
$\partial/\partial y\rightarrow ik_y$.  We also adopt the Lagrangian
time derivative operator
\begin{equation}
{\partial\over\partial t^\prime}={\partial\over\partial t}-q\Omega x
{\partial\over\partial y}.
\end{equation}

The explicit time-dependence of the radial wavenumber implies that a
Lagrangian treatment of the perturbation equations is much easier than
an Eulerian treatment.  We therefore work in terms of the Lagrangian
displacement vector $\bxi$, which to linear order is related to the
Eulerian velocity perturbation by
\begin{equation}
\delta{\bf v}={\partial{\bxi}\over\partial t^\prime}+q\Omega\xi_x\hat{\bf y}.
\end{equation}
This allows us to immediately solve the continuity and
flux-freezing equations to express the Eulerian density and magnetic field
perturbations in terms of the Lagrangian displacement \citep{new62}.  To
linear order,
\begin{equation}
\delta\rho=-i\rho k_x\xi_x-i\rho k_y\xi_y-\rho{\partial\xi_z\over\partial z}
-\xi_z {d\rho\over dz},
\end{equation}
\begin{equation}
\delta B_x=ik_yB\xi_x,
\end{equation}
\begin{equation}
\delta B_y=-\xi_z{dB\over dz}-B{\partial\xi_z\over\partial z}-iBk_x\xi_x,
\end{equation}
and
\begin{equation}
\delta B_z=ik_yB\xi_z.
\end{equation}

Eliminating the density and magnetic field perturbations, the linearized
momentum equations are then
\begin{equation}
{\partial^2\xi_x\over\partial t^{\prime2}}=-i{k_x\over\rho}\delta P-
(k_x^2+k_y^2)v_{\rm A}^2\xi_x+2q\Omega^2\xi_x+2\Omega
{\partial\xi_y\over\partial t^{\prime}}-i{k_xv_{\rm A}^2\over2H_{\rm mag}}\xi_z
+ik_xv_{\rm A}^2{\partial\xi_z\over\partial z},
\label{eqxix1}
\end{equation}
\begin{equation}
{\partial^2\xi_y\over\partial t^{\prime2}}=-i{k_y\over\rho}\delta P-2\Omega
{\partial\xi_x\over\partial t^\prime}-i{k_yv_{\rm A}^2\over2H_{\rm mag}}\xi_z,
\label{eqxiy1}
\end{equation}
and
\begin{eqnarray}
{\partial^2\xi_z\over\partial t^{\prime2}}&=&-{1\over\rho}
{\partial\delta P\over\partial z}-k_y^2v_{\rm A}^2\xi_z+
{v_{\rm A}^2\over(H^\prime_{\rm mag})^2}\xi_z-{3v_{\rm A}^2\over2H_{\rm mag}}
{\partial\xi_z\over\partial z}+v_{\rm A}^2{\partial^2\xi_z\over\partial z^2}
-i{k_xv_{\rm A}^2\over H_{\rm mag}}\xi_x\cr
&&+ik_xv_{\rm A}^2{\partial\xi_x\over\partial z}
+igk_x\xi_x+igk_y\xi_y+g{\partial\xi_z\over\partial z}-
{g\over H_\rho}\xi_z.
\label{eqxiz1}
\end{eqnarray}

The thermal pressure perturbation can be derived from the linearized energy
equation.  Assuming negligible gas pressure, and eliminating the flux
perturbation using the radiation diffusion equation, this is
\begin{equation}
3{\partial\delta P\over\partial t^\prime}+\delta v_z{dE\over dz}+
{4\over3}E\nabla\cdot\delta{\bf v}=-F{\partial\tilde{\delta\rho}\over
\partial z}+{c\over\kappa\rho H_\rho}{\partial\delta P\over\partial z}
+{c\over\kappa\rho}(k_x^2+k_y^2)\delta P+{c\over\kappa\rho}{\partial^2\delta P
\over\partial z^2}.
\end{equation}

Thus far, all of our equations are exact in the zero gas pressure limit.  We
now employ the WKB approximation in $z$, assuming that all perturbations
have a $z$-dependence of $\exp(ik_zz)$ and take all wavenumber components
to be large.  Because we are looking for instabilities with growth rates
that increase with wavenumber no faster than $k^{1/2}$, the time derivative
term in the energy equation is small.  Physically, radiative diffusion is
extremely fast at short wavelengths compared to our instability growth times,
i.e. we are in the rapid diffusion limit.  Eliminating the time derivative
here eliminates the damped diffusion mode of equation~(\ref{eqdiffmode}).
The remaining terms with the largest dependence on wavenumber then allow us
to approximate the thermal pressure perturbation as
\begin{equation}
\delta P\simeq{\kappa\rho\over ck^2}\left[Fk_z(k_x\xi_x+k_y\xi_y+k_z\xi_z)
-{4E\over3}\left(ik_x{\partial\xi_x\over\partial t^\prime}
+ik_y{\partial\xi_y\over\partial t^\prime}
+ik_z{\partial\xi_z\over\partial t^\prime}+ik_yq\Omega\xi_x\right)\right].
\end{equation}

The momentum equations can also be simplified by first noting that the
second order time derivative terms on the left hand sides of equations
(\ref{eqxix1}) and (\ref{eqxiz1}) are negligible given the $\sim k^2$ magnetic
force terms on the right hand sides.  Physically, eliminating these time
derivatives eliminates the fast and Alfv\'en modes of equations
(\ref{eqfastmode}) and (\ref{eqalfvenmode}).  Equations (\ref{eqxix1}) and
(\ref{eqxiz1}) can then be used to eliminate $\xi_x$ and $\xi_z$ from
equation (\ref{eqxiy1}).  The algebra is greatly simplified if one first
recognizes that in this limit of negligible gas pressure, slow modes have
fluid displacements that are predominately along the magnetic field
(in the $y$-direction).  In fact, $\xi_x$ and $\xi_z$ are one power of $k$
smaller than $\xi_y$.  Keeping only the lowest order terms as
$k\rightarrow\infty$, and assuming that magnetic gradients dominate
radiation pressure gradients in supporting the equilibrium (so that
$k_{\rm tran}H_{\rm mag}\gg1$ and the Parker regime generally exists in
the WKB limit for all wavenumber orientations), we finally obtain the
following differential equation:
\begin{equation}
{\partial^2\xi_y\over\partial t^{\prime2}}+{k_y^2\over k^2}{4E\kappa\over3c}
{\partial\xi_y\over\partial t^{\prime}}-{(k_x^2+k_y^2)v_{\rm A}^2\over4
H_{\rm mag}^2k^2}\xi_y+i{k_y^2k_z\over k^2}{\kappa F\over c}\xi_y=0.
\end{equation}
If there were no shear ($q=0$), so that the radial wavenumber
$k_x$ was independent of time, we could could assume a complex exponential
time-dependence $\xi_y\propto\exp(-i\omega t^\prime)$ and recover the
static equilibrium dispersion relation (\ref{eqdispparkerpbi}).

The Coriolis accelerations are simply too small to have affected the modes
at short wavelengths.  To order of magnitude, assuming a typical Parker
growth rate $\sim v_{\rm A}/H_{\rm mag}$, the Coriolis terms in equations
(\ref{eqxix1}) and (\ref{eqxiy1}) are negligible provided
$\Omega\ll kv_{\rm A}$.  But for a magnetically supported equilibrium,
$\Omega\sim v_{\rm A}/H_{\rm mag}$, so this is just equivalent to the WKB
condition $kH_{\rm mag}\gg1$.  In more detail, we require $|k_z|H_{\rm mag}\gg
k|k_y|/(|k_x|k_\perp)$, so that large radial wavenumbers better ensure
immunity of the Parker instability from Coriolis effects \citep{shu74}.
Provided $k_{\rm tran}H_{\rm mag}\gg1$, the
photon bubble regime will be even less affected by Coriolis forces.

\end{document}